\documentclass[11pt]{article}
\usepackage{graphicx}

\newcommand{\D}{\displaystyle}
\newcommand{\B}{\mathbf}
\newcommand{\p}{\partial}
\newcommand{\al}{\alpha}

\newcommand{\Bt}{\mbox{\boldmath$\tau$}}
\newcommand{\myref}[1]{(\ref{#1})}

\title{Removing the Stiffness of Elastic Force from the Immersed Boundary
Method for the 2D Stokes Equations}
\date{June 16, 2007}
\begin{document}
\author{Thomas Y. Hou\thanks{Applied and Comput. Math, Caltech, Pasadena, CA 91125. Email: hou@acm.caltech.edu.}
\and Zuoqiang Shi \thanks{Applied and Comput. Math, Caltech, Pasadena,
CA 91125 and Zhou Pei-Yuan Center for Applied Mathematics,
Tsinghua University, Beijing 100084, China. Email: shi@acm.caltech.edu.} }
\maketitle

\begin{abstract}
The Immersed Boundary method has evolved into one of the most useful
computational methods in studying fluid structure interaction. On the
other hand, the Immersed Boundary method is also known to suffer
from a severe timestep stability
restriction when using an explicit time discretization. In this paper,
we propose several efficient semi-implicit schemes to remove this stiffness
from the Immersed Boundary method
for the two-dimensional Stokes flow. First, we obtain a novel unconditionally
stable semi-implicit discretization for the immersed boundary problem.
Using this unconditionally stable discretization as a building block,
we derive several efficient semi-implicit schemes for the immersed
boundary problem by applying the Small Scale Decomposition to this
unconditionally stable discretization. Our stability analysis and
extensive numerical experiments show that our semi-implicit schemes
offer much better stability property than the explicit scheme. Unlike
other implicit or semi-implicit schemes proposed in the literature,
our semi-implicit schemes can be solved explicitly in the spectral
space. Thus the computational cost of our semi-implicit schemes is
comparable to that of an explicit scheme, but with a much better
stability property.
\end{abstract}

\section{Introduction}
The Immersed Boundary method was originally introduced by Peskin in
the 1970's to model the flow around heart valves. Now it has evolved
into a general useful method in studying the motion of one or more
massless, elastic surface immersed in an incompressible, viscous fluid,
particularly in biofluid dynamics problems where complex geometries
and immersed elastic membranes are present. The method has been
successfully applied to a variety of problems including blood flow
in the heart \cite{Peskin77,MPY82,MP83,MP85,PM89,MP89,MP01},
vibrations of the cochlear basilar membrane \cite{Beyer92,Givelberg97},
platelet aggregation during clotting \cite{Fogelson84,WF99},
aquatic locomotion \cite{FP88,Fauci90,HF02,ZP02,CCDF04},
flow with suspended particles \cite{FP88,SG98}, and
inset flight \cite{MP04,MP05},
We refer to \cite{Peskin02} for an extensive list of applications.

The Immersed Boundary method employs a uniform Eulerian grid over the
entire domain to describe the velocity field of the fluid and a
Lagrangian description for the immersed elastic structure. The force
generated by the elastic structure drives the fluid and the fluid
moves the elastic structure. This interaction is expressed in terms
of the spreading and interpolation operations by use of smoothed
Delta functions.

One of the main difficulties that the Immersed Boundary method
encounters is that it suffers from a severe timestep restriction
in order to keep the stability \cite{Peskin02,SW99,SW95}. This has
been the major
limitation of the Immersed Boundary method. This restriction is
typically much more severe than the one that would be imposed from
using an explicit discretization for the convection term in the
Navier-Stokes equation. The instability is known to arise from
large boundary force and small viscosity \cite{SW99}.
Much effort has been made to remove this restriction. Some implicit
and semi-implicit methods have been proposed in the literature
\cite{TP92,MP07,MP93}.
Despite of these efforts, the timestep restriction has not been
resolved satisfactorily. The computational cost of using an implicit
or semi-implicit scheme is still too high to be effective in a
practical computation. To date, almost all practical computations
using the immersed boundary method have been performed using
an explicit discretization.

In this paper, we develop several efficient semi-implicit schemes to
compute the motion of an elastic interface immersed in a two-dimensional,
incompressible Stokes flow. There are several important ingredients
in deriving our semi-implicit schemes. The first one is to use
the arclength and tangent angle formulation to describe the dynamics
of the immersed interface \cite{HLS94}. We remark that Ceniceros and Roma
have also used the arclength and tangent angle formulation to
alleviate the stiffness of the viscous
vortex sheet with surface tension in \cite{CR04}. The second one is to obtain
an unconditionally stable semi-implicit discretization of the immersed
boundary problem.
Throughout this paper, we use the term ``stability'' to mean that the
energy norm of the solution can be bounded in terms of the energy norm of
the initial data, which is a weaker result than proving that the difference
between two solutions in the energy norm can be bounded in terms of the
energy norm of their difference at time zero.
The third ingredient
is to perform Small Scale Decomposition to the unconditionally stable
discretization to obtain our efficient semi-implicit schemes. An
important feature of our small scale decomposition is that the leading
order term, which is to be discretized implicitly, can be expressed as
a convolution operator. This property enables us to solve for the
implicit solution explicitly using the Fourier transformation. Thus,
the computational cost of our semi-implicit schemes is comparable
to that of an explicit method. This offers a significant computational
saving in using the Immersed Boundary method.

The Small Scale Decomposition was first developed by Hou, Lowengrub
and Shelley \cite{HLS94,HLS97}. They applied this method to remove the
stiffness from interfacial flow with surface tension, which has
proved to be very successful. Due to the coupling between the elastic
boundary with the fluid, it is more difficult to remove the
stiffness induced by the elastic force in the Immersed Boundary
method. To remove the stiffness in the Immersed Boundary method,
we need to decouple the stiffness induced by the elastic force
from the fluid flow in such a way that the resulting semi-implicit
discretization is still unconditionally stable.
This is accomplished by using a semi-implicit discretization
 which preserves certain important solution structures which
exist at the continuous level. Without obtaining this unconditionally
stable semi-implicit discretization, a straightforward application
of the Small Scale Decomposition to the Immersed Boundary method
would not provide an efficient semi-implicit scheme with the desirable
stability property. Very recently, Newren et al. have obtained an
unconditionally
stable discretization for linear force in \cite{NFGK07}. However,
they did not perform Small Scale Decomposition to their unconditionally
stable discretization. As we will demonstrate in this paper, the
unconditionally stable semi-implicit discretization without using
the Small Scale Decomposition is still very expensive and the
gain over the explicit discretization is quite limited.

We develop several efficient semi-implicit schemes for both the steady
Stokes flow and the unsteady Stokes flow respectively. In both cases,
our semi-implicit schemes work very well. In the steady Stokes flow, we
also develop a fourth order semi-implicit scheme by using the integral
factor method. For the unsteady Stokes flow, we develop a second order
semi-implicit method by combining our Small Scale Decomposition with
a well known second order temporal discretization \cite{LP00,Peskin02}.
To illustrate
the stability properties of our semi-implicit schemes, we apply our
methods to several prototype problems and test our schemes for a
range of elastic coefficients and viscosity coefficients. Our
numerical results confirm that the semi-implicit schemes remove
the high order stability constraint induced by the elastic force.
In the case of unsteady Stokes equation, we also confirm the
second order accuracy of our semi-implicit scheme.

This paper is organized as follows. First, we review the classical
formulation of the Immersed Boundary method in Section 2. Then, we
introduce the arclength and tangent angle formulation in Section 3.
In Section 4, we describe the spatial discretization of the Immersed
Boundary method. In Section 5-6, we develop the numerical schemes for
steady Stokes flow and unsteady Stokes flow respectively.
The numerical results are presented in Section 7. Our numerical studies
will focus on the stability restriction and computational cost of
our methods. Some concluding remarks are given in Section 8.

\section{Review of the Immersed Boundary method}

For simplicity, we just consider a viscous incompressible fluid in
a two dimensional domain $\Omega$, containing an immersed massless
elastic boundary in the form of a closed simple curve $\Gamma$. The
configuration of the boundary is given in a parametric form:
$\B{X}(\alpha,t), 0\leq \alpha \leq L_b$,
$\B{X}(0,t)=\B{X}(L_b,t)$, $\alpha$ tracks a material point of the
boundary. We consider only the Stokes equations in this paper and
would neglect the convection term. Then the governing equations are
given as follows:
\begin{eqnarray}
\label{NS in review}
\rho \D\frac{\p\B{u}}{\p t}&=&-\nabla p + \mu \triangle
\mathbf{u} + \mathbf{f}(\mathbf{x},t)\;, \\
\label{incomp}
\nabla \cdot \mathbf{u} &=& 0 \;,\\
\label{eqa of boundary}
\displaystyle\frac{\partial \mathbf{X}}{\partial
t}(\alpha,t)&=&\mathbf{u(X}(\alpha,t),t) \;,
\end{eqnarray}
where
$\B{u}$ is the fluid velocity, $p$ is the pressure,
$\rho$ and $\mu$ are constant fluid density and viscosity,
$\B{f}(\mathbf{x},t)$ is the force density, which is not zero only on the
boundary and which is infinite there. The force density can be expressed
as below
\begin{eqnarray}
\label{force density}
\displaystyle\mathbf{f}(\mathbf{x},t)=\int_0^{L_b}\mathbf{F}(\alpha,t)\delta
(\mathbf{x-X}(\alpha,t))d\alpha ,
\end{eqnarray}
$\delta$ denotes the two-dimensional Dirac delta function and
\begin{eqnarray}
\displaystyle\mathbf{F}(\alpha,t)=\frac{\partial }{\partial
\alpha}(T\mbox{\boldmath$\tau$}) ,
\end{eqnarray}
\begin{eqnarray}
 T=T\left(\left|\frac{\partial \mathbf{X}}{\partial
\alpha}\right|\right).
\end{eqnarray}
The choice of function $T$ in this paper is computed by Hook's law
\begin{eqnarray}
T=S_b \left(\left|\frac{\partial \mathbf{X}}{\partial
\alpha}\right|-1\right) ,
\end{eqnarray}
where $S_b$ is the elastic coefficient of the boundary,
and $ \mbox{\boldmath$\tau$}$ is the unit tangent vector
along the boundary, which is defined as
\begin{eqnarray}
\mbox{\boldmath$\tau$}=\frac{\partial \mathbf{X}}{\partial
s}\Bigg/\left|\frac{\partial \mathbf{X}}{\partial s}\right|  .
\end{eqnarray}
This choice of force density has been used widely in the
literature in both computational and theoretical studies
\cite{JP01},\cite{RP01},\cite{TP92}.

We can rewrite \myref{eqa of boundary} in the following way:
\begin{eqnarray}
\label{eq of boundary}
\displaystyle\frac{\partial \mathbf{X}}{\partial
t}(\alpha,t)&=&\int_\Omega\B{u}(\B{x},t)\delta(\B{x}-\B{X}(\alpha,t))d\B{x}.
\end{eqnarray}
Next, we introduce the spreading and interpolation operations.
The spreading and interpolation operators are defined as follows:
\begin{eqnarray}
L(\B{X})(g(\alpha))(\B{x})&=&\int_\Gamma g(\alpha)\delta(\B{x}-\B{X}(\alpha,t))d\alpha ,\\
L^*(\B{X})(u(\B{x}))(\alpha)&=&\int_\Omega u(\B{x})\delta(\B{x}-\B{X}(\alpha,t))d\B{x} \;.
\end{eqnarray}
It is easy to show that $L$ and $L^*$ are adjoint operators:
\begin{eqnarray}
&&<\B{u}(\B{x}),L(\B{X})(g(\alpha))>_\Omega\nonumber\\
&=&\int_\Omega \B{u}(\B{x}) \left(\int_\Gamma g(\alpha)\delta(\B{x}-\B{X}(\alpha,t))d\alpha\right) d\B{x}\nonumber\\
&=&\int_\Omega  \int_\Gamma \B{u}(\B{x})g(\alpha)\delta(\B{x}-\B{X}(\alpha,t))d\alpha d\B{x}\nonumber\\
&=&  \int_\Gamma \int_\Omega\B{u}(\B{x})g(\alpha)\delta(\B{x}-\B{X}(\alpha,t)) d\B{x}d\alpha\nonumber\\
&=&  \int_\Gamma g(\alpha)\left(\int_\Omega\B{u}(\B{x})\delta(\B{x}-\B{X}(\alpha,t)) d\B{x}\right)d\alpha\nonumber\\
&=&<L^*(\B{X})(u(\B{x})),g(\alpha)>_\Gamma \;,
\end{eqnarray}
where the inner product are defined as follows:
\begin{eqnarray}
<u,v>_\Omega&=&\int_\Omega u(\B{x})v(\B{x})d\B{x} ,\\
<f,g>_\Gamma&=&\int_\Gamma f(\al)g(\al)d\al .
\end{eqnarray}

Equations \myref{NS in review},\myref{incomp} are the
familiar Stokes equations of viscous incompressible fluid.
Equations \myref{eqa of boundary},\myref{force density} represent the
interaction of the fluid and the elastic boundary. The elastic boundary
applies the force to the fluid, the fluid carries the immersed
boundary, and the force density is determined by the configuration of
the boundary.

\section{The arclength-tangent angle formulation}

In studying the evolution of a curve, it is useful to represent the curve by
its tangent angle $\theta$ and local arclength derivative $s_\al$.
Previously, Hou, Lowengrub and Shelley \cite{HLS94} exploited this
formulation and combined it with a so-called "Small Scale Decomposition"
reformulation to remove the stiffness induced by surface tension.

Consider the evolution of a simply closed curve $\Gamma$ with known normal
and tangent velocity fields, $U,V$. Assume the curve is represented by
$\B{X}(\al,t), \al\in [0,L_b]$. We define the arclength derivative,
$s_\alpha$, and the tangent vector, $\theta$, as follows
\begin{eqnarray}
&&s_\alpha (\alpha,t) = | \B{X}_\alpha (\alpha, t) |, \\
&&(x_\al(\al,t),y_\al(\al,t)) = s_\al(\al,t)(\cos\theta(\al,t),\sin\theta(\al,t)).
\label{tangent vector}
\end{eqnarray}
The closed curve $\Gamma$ evolves according to
\begin{eqnarray}
\frac{\p \B{X}}{\p t}=\B{u}(\B{X},t)=U\B{n}+V\mbox{\boldmath$\tau$},
\end{eqnarray}
where $\mbox{\boldmath{$\tau$}}$ and $\B{n}$ are the unit tangent and
normal vectors of the curve respectively.
According to the Frenet formula, we have
 $\frac{\p \mbox{\boldmath$\tau$}}{\p s}=k\B{n}, \frac{\p \B{n}}{\p s}=-k\mbox{\boldmath$\tau$}$, here $s$ is the arclength variable.
It is easy to see that $s_\al$ and $\theta$ satisfy the following
evolution equations \cite{HLS94}:
\begin{eqnarray}
\label{eq of sa}
(s_{\alpha})_t&=&V_\alpha-\theta_\alpha U ,\\
\label{eq of theta}
\theta_t&=&\frac{U_\alpha}{s_\alpha}+\frac{V\theta_\alpha}{s_\alpha} .
\end{eqnarray}
Given $s_\al$ and $\theta$, the curve $\Gamma$ can be reconstructed up
to a translation by integrating \myref{tangent vector}. However, we also
need a point on the boundary to provide the constant of integration.

Using the $s_\al-\theta$ formulation, we can reformulate the
immersed boundary problem as follows:
\begin{eqnarray}
\rho \frac{\p\B{u}}{\p t}&=&-\nabla p + \mu \triangle \mathbf{u} + L(\B{X})\left(\mathbf{F}(s_\al,\theta)\right), \\
\nabla \cdot \mathbf{u} &=& 0 \\U&=&L^*(\B{X})(\B{u}(\B{x}))\cdot \B{n},\\
V&=&L^*(\B{X})(\B{u}(\B{x}))\cdot \mbox{\boldmath$\tau$},\\
(s_{\alpha})_t&=&V_\alpha-\theta_\alpha U,\\
\theta_t&=&\frac{U_\alpha}{s_\alpha}+\frac{V\theta_\alpha}{s_\alpha} ,
\end{eqnarray}
where
\begin{eqnarray}
\displaystyle\mathbf{F}(s_\alpha,\theta)=\frac{\partial }{\partial
\alpha}(T\mbox{\boldmath$\tau$})=S_b \left(s_{\al,\al}\mbox{\boldmath$\tau$}+(s_\al-1)\theta_\al\B{n}\right) .
\end{eqnarray}

\section{Spatial Discretization}

We use the spectral method to discretize both the Stokes
equations and the immersed boundary equations in space
since we work on periodic domains. We first discuss the
discretization of the Stokes equations in a regular
$N\times N$ Cartesian grid with a uniform meshsize $h$.
Let $x_j = jh$ and $y_j=jh$.
The discrete Fourier transform and inverse Fourier transform
are defined as follows:
\begin{eqnarray}
\mathcal{F}_{h,x}(\phi)(k,y)&=&\frac{1}{N}\sum_{j=0}^{N-1}
\phi(x_j,y)e^{-ikx_j}=\widehat{\phi}(k,y), \quad -N/2+1 \leq k \leq N/2,\\
\mathcal{F}_{h,y}(\phi)(x,k)&=&\frac{1}{N}\sum_{j=0}^{N-1}
\phi(x,y_j)e^{-iky_j}=\widehat{\phi}(x,k), \quad -N/2+1 \leq k \leq N/2,\\
\mathcal{F}^{-1}_{h,x}(\widehat{\phi})(x_j,y)&=&\sum_{k=-N/2+1}^{N/2}
\widehat{\phi}(k,y)e^{ikx_j}=\phi(x_j,y), \quad 0 \leq j \leq N-1,\\
\mathcal{F}^{-1}_{h,y}(\widehat{\phi})(x,y_j)&=&\sum_{k=-N/2+1}^{N/2}
\widehat{\phi}(x,k)e^{iky_j}=\phi(x,y_j), \quad 0 \leq j \leq N-1.
\end{eqnarray}
Now we introduce the discrete differential operator using the
discrete Fourier transform defined above. For a function $\phi(x,y)$
defined in the fluid domain $\Omega$, we approximate its spatial
derivatives as follows:
\begin{eqnarray}
\left(D_{h,x} \phi\right)(x,y)=\mathcal{F}^{-1}_{h,x} \left(ik\left(\mathcal{F}_{h,x} \phi\right)(k,y)\right),\\
\left(D_{h,y} \phi\right)(x,y)=\mathcal{F}^{-1}_{h,y} \left(ik\left(\mathcal{F}_{h,y} \phi\right)(x,k)\right) .
\end{eqnarray}
Denote $\nabla_h=(D_{h,x},D_{h,y})$. The differential operators are discretized
in terms of $\B{D}_h$:
\begin{eqnarray}
\nabla p&\rightarrow&  \nabla_h p, \\
\nabla \cdot \B{u}&\rightarrow& \nabla_h \cdot \B{u},\\
\nabla^2 u &\rightarrow& \nabla_h\cdot \nabla_h\, u\equiv \nabla^2_h u .
\end{eqnarray}

Next, we describe the discretization of the immersed boundary.
We employ a Lagrangian grid with grid space $\Delta \al$. The number of
grid points along the boundary is $N_b$. For a function $\psi(\al)$ defined
on the interface $\Gamma$, we define the discrete Fourier transform and its
inverse as follows:
\begin{eqnarray}
\mathcal{F}_{\Delta \al}(\psi)(k)&=&\frac{1}{N_b}\sum_{j=0}^{N_b-1}\phi(\al_j)e^{-ik\al_j}=\widehat{\psi}(k), \quad \al_j=j\Delta \al ,\\
\mathcal{F}^{-1}_{\Delta \al}(\widehat{\psi})(\al_j)&=&
\sum_{k=-\frac{N_b}{2}+1}^{\frac{N_b}{2}}\widehat{\psi}(k)e^{ik\al_j}=
\psi(\al_j).
\end{eqnarray}
When the interface is a closed curve, we can approximate the derivative
operator along the interface by the spectral derivative:
\begin{eqnarray}
\left(D_{\Delta \al} \psi\right)(\al)=\mathcal{F}^{-1}_{\Delta \al} \left(ik\left(\mathcal{F}_{\Delta \al} \phi\right)(k)\right).
\end{eqnarray}
When the solution is not periodic, we can also use a finite difference method
to discretize the derivative, we refer to \cite{Peskin02} for more details.

Now we discuss the discretization of the spreading and interpolation
operators. These two operators both involve the use of a discrete
delta function. The discrete delta function we use is
introduced by Peskin in \cite{Peskin02}:
\begin{eqnarray}
\delta_h(x,y)=\frac{1}{h^2}\phi\left(\frac{x}{h}\right)\phi
\left(\frac{y}{h}\right),
\end{eqnarray}
and
\begin{eqnarray}
\phi(r)=\left\{
\begin{array}{cc}
\frac{1}{8}\left(3-2|r|+\sqrt{1+4|r|-4r^2}\right),& |r|\le 1,\\
\frac{1}{8}\left(5-2|r|-\sqrt{-7+12|r|-4r^2}\right),& 1\le |r|\le 2,\\
0, & |r|>2 .
\end{array}
\right.
\end{eqnarray}
Using the above discrete delta function, we can discretize the spreading and interpolation operator as follows
\begin{eqnarray}
\label{dis of spreading}
L_h(\B{X})(g(\alpha))(\B{x})&=&\sum_{\al \in \mathcal{G}_\Gamma} g(\alpha)\delta_h(\B{x}-\B{X}(\alpha,t))\Delta\al ,\\
\label{dis of inter}
L^*_h(\B{X})(u(\B{x}))(\alpha)&=&\sum_{\B{x}\in \mathcal{G}_\Omega} u(\B{x})\delta_h(\B{x}-\B{X}(\alpha,t))h^2 .
\end{eqnarray}
The summation above is over grid points in $\Gamma$ in \myref{dis of spreading} and over grid points in $\Omega$ in
\myref{dis of inter}.
Operator $L_h$ and $L^*_h$ are still adjoint using the following
discrete inner product:
\begin{eqnarray}
\label{inner product gamma}
<f,g>_{\Gamma_h}=\sum_{\al \in \mathcal{G}_\Gamma} f(\al)g(\al)\Delta \al,\\
\label{inner product omega}
<u,v>_{\Omega_h}=\sum_{\B{x} \in
\mathcal{G}_\Omega} u(\B{x})v(\B{x})h^2.
\end{eqnarray}
Using the inner product defined above, we have:
\begin{eqnarray}
&&<u(\B{x}), L(\B{X})(g(\al))>_{\Omega_h}\nonumber\\
&=&\sum_{\B{x} \in \mathcal{G}_\Omega} u(\B{x})L(\B{X})(g(\al))h^2\nonumber\\
&=&\sum_{\B{x} \in \mathcal{G}_\Omega} u(\B{x})h^2\sum_{\al \in \mathcal{G}_\Gamma} g(\alpha)\delta_h(\B{x}-\B{X}(\alpha,t))\Delta\al\nonumber\\
&=&\sum_{\B{x} \in \mathcal{G}_\Gamma}g(\alpha)\Delta\al \sum_{\al \in \mathcal{G}_\Omega} u(\B{x})\delta_h(\B{x}-\B{X}(\alpha,t))h^2\nonumber\\
&=&<L^*_h(\B{X})(u(\B{x})), g(\al)>_{\Gamma_h} .
\end{eqnarray}
As we will see later, this discrete self-adjoint property is crucial
in obtaining our unconditional stable semi-discrete scheme for the
immersed boundary problem.

\section{Steady Stokes flow}
\subsection{Formulation}
For simplicity, we study the steady Stokes flow first.
The governing equations for the steady Stokes flow are given as follows:
\begin{eqnarray}
\label{steady Stokes}
0&=&-\nabla p + \mu \triangle \mathbf{u} + L(\B{X})\left(\mathbf{F}(s_\al, \theta)\right) ,\\
\label{incomp of steady}
\nabla \cdot \mathbf{u} &=& 0, \\
U&=&\B{u(X}(\alpha,t),t)\cdot \B{n},\\
V&=&\mathbf{u(X}(\alpha,t),t)\cdot \mbox{\boldmath$\tau$},\\
(s_{\alpha})_t&=&V_\alpha-\theta_\alpha U,\\
\theta_t&=&\frac{U_\alpha}{s_\alpha}+\frac{V\theta_\alpha}{s_\alpha}.
\end{eqnarray}

In this simple case, we can use a boundary integral method for the
two dimension Stokes flow (page 60 of \cite{Pozrikidis92}) to solve
equations \myref{steady Stokes}-\myref{incomp of steady} to get the
velocity on the boundary:
\begin{eqnarray}
\label{u of steady}u(\mathbf{X}(\alpha,t))&=&\frac{1}{4\pi \mu}\int_\Gamma\left( -(\ln r)
F_1(\alpha',t)+\frac{r_1^2}{r^2}F_1(\alpha',t)+\frac{r_1r_2}{r^2}F_2(\alpha',t)\right)d\alpha',\\
\label{v of steady}
v(\mathbf{X}(\alpha,t))&=&\frac{1}{4\pi \mu}\int_\Gamma\left( -(\ln r)
F_2(\alpha',t)+\frac{r_2^2}{r^2}F_2(\alpha',t)+\frac{r_1r_2}{r^2}F_1(\alpha',t)\right)d\alpha' ,
\end{eqnarray}
where $ r=|\B{r}| $ and
\begin{eqnarray}
\B{r}=(r_1,r_2)=\B{X}(\alpha,t)-\B{X}(\alpha',t),\quad \B{F}=(F_1,F_2),\quad \B{u}=(u,v).
\end{eqnarray}

\subsection{Small Scale Decomposition}

As we can see from \myref{u of steady}-\myref{v of steady}, the velocity
field can be expressed as a singular integral with a kernel $\ln(r)$.
However, the singular velocity integral is nonlinear and nonlocal. It is
difficult to solve for the implicit solution if we treat the
velocity integral fully implicitly. The main idea of the
Small Scale Decomposition technique introduced in \cite{HLS94}
is to decompose the singular velocity integral into the sum of a
linear singular operator which is a convolution operator independent of
time $t$ and the configuration of the curve, and a remainder operator
which is regular. Since the remaining operator, which is nonlinear
and nonlocal, is regular, the simplified convolution integral operator
captures accurately the high frequency spectral property of the original
velocity integral. Thus, if we treat only the leading order convolution
operator implicitly, but keep the regular remainder operator explicitly,
we can effectively remove the stiffness of the original velocity
field which comes mainly from the high frequency modes of the solution.
In this subsection, we will show how to perform such Small Scale
Decomposition for the Immersed Boundary method applied to
the Stokes flow.

Observe that in the integral representation of the velocity field,
\myref{u of steady}-\myref{v of steady}, the only singular part of
the kernel is $\ln(r)$. The other part of the kernel is smooth.
Thus to the leading order contribution of the velocity field
can be expressed as follows:
\begin{eqnarray}
\B{u}(\mathbf{X}(\alpha,t))\sim\frac{1}{4\pi \mu}\int_\Gamma -(\ln r)
\B{F}(\alpha',t)d\al',
\end{eqnarray}
\begin{eqnarray}
\label{lead V}
V&=&\mathbf{u(X}(\alpha,t),t)\cdot \mbox{\boldmath$\tau$}(\al)\\
&\sim&\frac{1}{4\pi \mu}\int_\Gamma -(\ln r)
\B{F}(\alpha',t)\cdot \mbox{\boldmath$\tau$}(\al)d\al'\nonumber\\
&=&\frac{S_b}{4\pi \mu}\int_\Gamma -(\ln r)
\left(s_{\al,\al'}\mbox{\boldmath$\tau$}(\al')+(s_\al-1)\theta_{\al'}\B{n}(\al')\right)\cdot \mbox{\boldmath$\tau$}(\al)d\al' .
\end{eqnarray}
Next, we perform a Taylor expansion for $r$,
$\mbox{\boldmath$\tau$}(\al')\cdot \mbox{\boldmath$\tau$}(\al)$
and $\B{n}(\al')\cdot \mbox{\boldmath$\tau$}(\al)$ as a function of
$\al'$ around $\al$. By keeping only the leading order term, we have
\begin{eqnarray}
r\sim s_\al(\al)|\al-\al'|,\\
\mbox{\boldmath$\tau$}(\al')\cdot \mbox{\boldmath$\tau$}(\al)\sim 1,\\
\B{n}(\al')\cdot \mbox{\boldmath$\tau$}(\al)\sim 0.
\end{eqnarray}
Substituting the above Taylor expansions to \myref{lead V}, we get
\begin{eqnarray}
V&\sim&\frac{S_b}{4\pi\mu}\int\ln\left(s_\al(\al)|\al-\al'|\right)s_{\al,\al'}d\alpha'.
\end{eqnarray}
Integrating by part, we obtain
\begin{eqnarray}
V&\sim&\frac{S_b}{4\pi\mu}\int\frac{s_{\alpha'}}{\alpha'-\alpha}d\alpha'=-\frac{S_b}{4\mu}\mathcal{H}[s_\al],
\end{eqnarray}
where $\mathcal{H}$ is the Hilbert transform
\begin{eqnarray}
\mathcal{H}[f](\al)=\frac{1}{\pi}\int_{-\infty}^{+\infty}\frac{f(\al')}{\al-\al'}d\al'.
\end{eqnarray}
Using the same method, we can get the leading order contributions
of $U$ and $U_\al$ as follows:
\begin{eqnarray}U&\sim&\frac{S_b}{4\pi\mu}\int-\ln
|\alpha-\alpha'|(s_{\alpha'}-1)\theta_{\alpha'}d\alpha'\\
U_\alpha&\sim&\frac{S_b}{4\pi\mu}\int\frac{(s_{\alpha'}-1)\theta_{\alpha'}}{\alpha'-\alpha}d\alpha'=-\frac{S_b}{4\mu}\mathcal{H}[(s_{\alpha}-1)\theta_{\alpha}].
\end{eqnarray}

Note that if $f$ is a smooth function, then the commutator
$[\mathcal{H},f]u \equiv \mathcal{H}(fu) - f\mathcal{H}(u)$ is a smoothing
operator for $u$. Thus we can factor a smooth function from the
Hilbert transform without changing its leading order spectral property.
Suppose that $s_\al$ is smooth, then we obtain to the leading order that
\begin{eqnarray}U_\alpha&\sim&-\frac{S_b}{4\mu}(s_{\alpha}-1)\mathcal{H}[\theta_{\alpha}].
\end{eqnarray}

Applying the same analysis to the Eqs
\myref{eq of steady s}-\myref{eq of steady theta} gives
\begin{eqnarray}
\label{leading eq of sa}
&&(s_{\al})_t=-\frac{S_b}{4\mu}\mathcal{H}[D_{\Delta \alpha}s_{\al}]
+\left(D_{\Delta \alpha}V-
D_{\Delta \alpha}\theta U+\frac{S_b}{4\mu}\mathcal{H}[D_{\Delta \alpha}
s_{\al}]\right),\\
\label{leading eq of theta}
&&\theta_t=-\frac{S_b}{4\mu}\left(1-\frac{1}{s_\al}\right)\mathcal{H}[
D_{\Delta \alpha}\theta]+\left(\frac{D_{\Delta \alpha}U}{s_\alpha}+
\frac{V D_{\Delta \alpha}\theta}{s_\alpha}+\frac{S_b}{4\mu}
\left(1-\frac{1}{s_\al}\right)\mathcal{H}[D_{\Delta \alpha}\theta]\right).
\end{eqnarray}
Note that the leading order operator is linear. This suggests a natural
semi-implicit discretization of the immersed boundary problem.

Since we are dealing with a closed immersed boundary, it is natural to
work in the Fourier space. Furthermore, the Hilbert operator has a very
simple kernel under the Fourier transformation. Notice that $\theta$ is not
a periodic function of $\al$. Its value increases $2\pi$
every time $\al$ increases $L_b$. Nevertheless, if we let
\begin{eqnarray}
\label{periodic theta}
\theta(\al,t)=\frac{2\pi}{L_b}\al+\phi(\al,t), \;\al\in [0,L_b],
\end{eqnarray}
then $\phi$ is periodic. It is more convenient to work with $\phi$
than $\theta$. Substituting \myref{periodic theta} into
\myref{leading eq of theta} and taking the Fourier transform on
both sides of \myref{leading eq of sa},\myref{leading eq of theta},
we obtain
\begin{eqnarray}
\label{ft of eq of sa}
\hat{s}_{\al,t}&=&-\frac{S_b}{4\mu}|k|\hat{s}_\al+\left[\mathcal{F}
\left(D_{\Delta \alpha}V- D_{\Delta \alpha}\theta U\right)+
\frac{S_b}{4\mu}|k|\hat{s}_\al\right],\\
\label{ft of eq of theta}
\hat{\phi}_t&=&-\frac{S_b}{4\mu}\gamma|k|\hat{\phi}
+\left[\mathcal{F}\left(\frac{D_{\Delta \alpha}U}{s_\alpha}
+\frac{V D_{\Delta \alpha}\theta}{s_\alpha}\right)
+\frac{S_b}{4\mu}\gamma|k|\hat{\phi}\right],
\end{eqnarray}
where $\D\gamma=\max_\al\left(1-\frac{1}{s_\al}\right)$. We have also
used the fact that $\widehat{\mathcal{H}}_k =-i\;\mbox{sgn}(k)$ with
$\mbox{sgn}(k)$ being the signature function.
The first term on the right hand side captures the leading order
high frequency contribution of the terms from the right hand side.
An important property of this leading order term is that it is linear
in $\hat{s}_\al$ and $\hat{\theta}$ and has constant coefficient in
space. This provides a straightforward application of the
implicit time discretization.

Since our small scale decomposition is exact near the equilibrium, we
can use this result to get the stability constraint of the explicit
scheme by using a frozen coefficient analysis. The stability constraint
is given by
\begin{eqnarray}
\Delta t <C\frac{\mu}{S_b}\frac{h}{\gamma}.
\end{eqnarray}
As we can see, the time step needs to be very small if $S_b$ is
large and $\mu$ is small. For example, if $S_b = 100$, and $\mu
=10^{-2}$, then the stability would require that $\Delta t \leq C
10^{-4} h$.

\subsection{Semi-implicit schemes}

Based on the small scale decomposition presented in the previous
subsection, we propose two types of semi-implicit schemes in this
section. The first implicit time discretization uses the backward Euler
method to discretize the leading order term while keeping the
lower order term explicit. This gives rise to the following
semi-implicit scheme:
\begin{eqnarray}
\frac{\hat{s}_{\al}^{n+1}-\hat{s}_{\al}^{n}}{\Delta t}&=&
-\frac{S_b}{4\mu}|k|\hat{s}_\al^{n+1}+\left[\mathcal{F}\left(
D_{\Delta \alpha}V^n -D_{\Delta \alpha}\theta^n U^n\right)+\frac{S_b}{4\mu}|k|\hat{s}_\al^n\right],\\
\frac{\hat{\phi}^{n+1}-\hat{\phi}^{n}}{\Delta t}&=&-\frac{S_b}{4\mu}\gamma|k|\hat{\phi}^{n+1}+\left[\mathcal{F}\left(\frac{D_{\Delta \alpha}U^n}
{s_\alpha^{n+1}}+\frac{V^nD_{\Delta \alpha}\theta^n}{s_\al^{n+1}}
\right)+\frac{S_b}{4\mu}\gamma|k|\hat{\phi}^n\right].
\end{eqnarray}
We call the above discretization the semi-implicit method. Near
equilibrium, the stability constraint of this numerical method is
$\Delta t<C(S_b,\mu)$, independent of the meshsize $h$. Since the
small scale decomposition only captures the leading order
contribution from the high frequency components, this method can not
eliminate the effect of $S_b$ and $\mu$ completely. The coefficients
$S_b$ and $\mu$ can still affect the time stability through the low
frequency components of the solution, which comes from the second
term of the right hand side. In order to obtain a semi-implicit
discretization with better stability property, we can incorporate
the low frequency contribution from the second term in our implicit
discretization. This scheme can be found in the appendix $\B{A}$. We
call it the semi-implicit scheme of second kind.

The accuracy of the semi-implicit schemes presented above is just
first order. In order to get a high order time discretization, we can
use the integral factor method. The integral factor method factors
out the leading order linear term prior to time discretization. They
usually provide stable and high order time integration methods for stiff
problems. To use the integral factor method, we rewrite
\myref{ft of eq of sa},\myref{ft of eq of theta} as
\begin{eqnarray}
\frac{\p}{\p t}\left(e^{\eta t}\hat{s}_\al\right)&=&\exp\left(\frac{S_b}{4\mu}|k|t\right)P(\hat{s}_\al,\hat{\phi}),\\
\frac{\p}{\p t}\left(e^{\xi t}\hat{\phi}\right)&=&\exp\left(\frac{S_b}{4\mu}\gamma|k|t\right)Q(\hat{s}_\al,\hat{\phi}),
\end{eqnarray}
where
\begin{eqnarray}
\eta&=&\frac{S_b}{4\mu}|k|, \quad
\xi=\frac{S_b}{4\mu}\gamma|k|,\\
P(\hat{s}_\al,\hat{\phi})&=&\mathcal{F}\left(V_\alpha-\theta_\alpha U\right)+\frac{S_b}{4\mu}|k|\hat{s}_\al,\\
Q(\hat{s}_\al,\hat{\phi})&=&\mathcal{F}\left(\frac{U_\alpha}{s_\alpha}+\frac{V\theta_\alpha}{s_\alpha}\right)+\frac{S_b}{4\mu}\gamma|k|\hat{\phi}.
\end{eqnarray}
Now it is straightforward to discretize this system to high order.
In particular, we can apply the classical fourth order Runge-Kutta
method to discretize the above system to obtain a fourth order
semi-implicit scheme.

We remark that although the fourth order semi-implicit scheme based on
the integral factor approach is much more accurate than the first order
semi-implicit discretization, the stability of the fourth order method
is weaker than the first semi-implicit scheme based on the backward
Euler discretization. The fact that the higher order discretization
gives a weaker stability property is a phenomenon which has been
observed for almost all time integration methods. It is not a
restriction of our semi-implicit schemes for the immersed boundary problem.

The semi-implicit schemes we describe above only update the
$\theta$ and $s_\al$ variables. We also need to reconstruct the boundary
at $t^{n+1}$ from $\theta^{n+1}$ and $s_\al^{n+1}$. For this purpose, we
need to update a reference point of the boundary. This will be done
by using an explicit time integration method. The simplest one is
the forward Euler method:
\begin{eqnarray}
\label{x ref updation}
x^{n+1}(0)&=&x^{n}(0)+\Delta t\left(V^n\cos(\theta^n(0))-U^n\sin(\theta^n(0))\right),\\
\label{y ref updation}
y^{n+1}(0)&=&y^{n}(0)+\Delta t\left(V^n\sin(\theta^n(0))+U^n\cos(\theta^n(0))\right),
\end{eqnarray}
where $U$ and $V$ are evaluated at the reference point. A higher order
integration method can be also used. In the explicit update of the
reference point, we can use the values of $U$ and $V$ obtained using the
semi-implicit discretization from the previous time steps to extrapolate
the values of $U$ and $V$ in the intermediate time steps in our explicit
update of the reference point. Once we have updated the reference point,
we can obtain the configuration of the boundary $(x,y)$ from
$(s_\al,\theta)$ by integrating \myref{tangent vector}
\begin{eqnarray}
x^{n+1}(\al)&=&x^{n+1}(0)+\int_0^\al s_\al^{n+1}(\al') \cos(\theta^{n+1}(\al'))d\al',\\
y^{n+1}(\al)&=&y^{n+1}(0)+\int_0^\al s_\al^{n+1}(\al') \sin(\theta^{n+1}(\al'))d\al'.
\end{eqnarray}

We can use more than one reference point, then average them to get
the last configuration. This can improve the stability constraint
significantly. Actually, in our computation, we use two reference
points $X(0), X(N_b/2)$, then take the average to determine the position
of the interface at next time step. Since we update only two
reference points, the extra cost in updating the reference point is
small compared with the overall computational cost.

\section{Unsteady Stokes flow}
\subsection{Formulation}
In this section, we will extend the semi-implicit discretization
developed for the steady Stokes flow to the unsteady Stokes flow.
The governing equations of the immersed boundary method for
the unsteady Stokes flow are as follows:
\begin{eqnarray}
\label{unsteady Stokes eq}
\rho\frac{\p \B{u}}{\p t}&=&-\nabla p + \mu \triangle \mathbf{u} +L(\B{X})\left(\mathbf{F}(s_\al, \theta)\right) ,\\
\label{unsteady incomp}
\nabla \cdot \mathbf{u} &=& 0 \\U&=&\B{u(X}(\alpha,t),t)\cdot \B{n},\\
V&=&\mathbf{u(X}(\alpha,t),t)\cdot \mbox{\boldmath$\tau$},\\
s_{\alpha t}&=&V_\alpha-\theta_\alpha U,\\
\theta_t&=&\frac{U_\alpha}{s_\alpha}+\frac{V\theta_\alpha}{s_\alpha}.
\end{eqnarray}

It is much more difficult to solve the fluid velocity $\B{u}$ analytically
from \myref{unsteady Stokes eq}-\myref{unsteady incomp}. As for the
steady Stokes flow, we will first derive an unconditionally stable time
discretization which will be given in next section
 and then apply the Small Scale Decomposition to the
unconditionally stable time discretization to obtain our efficient
semi-implicit schemes.

\subsection{An unconditionally stable semi-implicit discretization}

In this section, we will describe our unconditionally stable
semi-implicit discretization of the Immersed Boundary method
for the incompressible unsteady Stokes equations and prove its unconditional
stability in the sense of total energy is non-increasing.

The unconditionally stable semi-implicit discretization is consisted of two steps.
In the first step, we update $s_\al, \B{u}$ from
$t^n$ to $t^{n+1}$, then we get $\theta^{n+1}$ in the second step.

\vspace{0.1in}
{\bf Step 1}: Update of $\B{u}^{n+1}$ and $s_\alpha^{n+1}$.
\begin{eqnarray}
\label{disc unsteady Stokes eq}
\rho\frac{\B{u}^{n+1}-\B{u}^n}{\Delta t}&=&-\nabla_h p^{n+1}+\mu \nabla_h^2 \B{u}^{n+1}+L_{h,n}\left(\B{F}(s_\alpha^{n+1},\theta^{n}; \Bt^n,\B{n}^n)\right),\\
\label{disc p}
\nabla_h^2 p^{n+1}&=&\nabla_h\cdot L_{h,n}\left(\B{F}(s_\alpha^{n+1},\theta^{n}; \Bt^n,\B{n}^n)\right), \\
V^{n+1}&=&L^*_{h,n}(\B{u}^{n+1})\cdot \Bt^{n}\\U^{n+1}&=&L^*_{h,n}(\B{u}^{n+1})\cdot \B{n}^{n},\\
\label{stable disc sa eq}
\frac{s_\alpha^{n+1}-s_\alpha^{n}}{\Delta t}&=&D_{\Delta \alpha}V^{n+1}
- D_{\Delta \alpha} \theta^{n}U^{n+1},
\end{eqnarray}
where $\Bt^{n} = (\cos(\theta^n),\sin(\theta^n))$,
$\B{n}^{n} = (-\sin(\theta^n), \cos(\theta^n))$,
 $L_{h,n}=L_h(\B{X}^n), L^*_{h,n}=L_h^*(\B{X}^n)$,
$\nabla_h$ and $D_{\Delta \alpha}$ are discrete derivative
operators for the Eulerian grid and the Lagrangian grid
respectively, and
\begin{eqnarray}
\label{F-eqn}
\B{F}(s_\alpha^{n+1},\theta^{n}; \Bt^n,\B{n}^n) &=&
S_b \left(D_{\Delta \alpha}s_\alpha^{n+1}\Bt^{n}
+(s_\alpha^{n+1}-1)D_{\Delta \alpha}\theta^{n} \B{n}^n\right).
\end{eqnarray}

\vspace{0.1in}
{\bf Step 2}: Update of $\theta^{n+1}$.
After we have obtained ${\bf u}^{n+1}$, $p^{n+1}$ and
$s_\alpha^{n+1}$, we update $\theta$ at $t^{n+1}$ using the following
semi-implicit scheme:
\begin{eqnarray}
\label{disc vel theta}
\rho\frac{\B{\overline{u}}^{\;n+1}-\B{u}^{n}}{\Delta t}&=&
-\nabla_h \overline{p}^{\;n+1}+\mu \nabla_h^2 \overline{\B{u}}^{\;n+1}
+L_{h,n}\left(\B{F}(s_\alpha^{n+1},\theta^{n+1}; \Bt^n,\B{n}^n)\right), \\
\label{disc p theta}
\nabla_h^2 \;\overline{p}^{\;n+1}&=&\nabla_h\cdot
L_{h,n}\left(\B{F}(s_\alpha^{n+1},\theta^{n+1}; \Bt^n,\B{n}^n)\right)
, \\
\overline{V}^{\;n+1}&=&L^*_{h,n}(\overline{\B{u}}^{\;n+1})\cdot \Bt^{n} \\
\overline{U}^{\;n+1}&=& L^*_{h,n}(\overline{\B{u}}^{\;n+1})\cdot \B{n}^{n},\\
\label{stable disc theta eq}
\frac{\theta^{n+1}-\theta^{n}}{\Delta t}&=&\frac{1}{s_\alpha^{n+1}}
\left(D_{\Delta \alpha}\overline{U}^{\;n+1}+D_{\Delta \alpha}\theta^{n}\overline{V}^{\;n+1}\right).
\end{eqnarray}
where
\begin{eqnarray}
\label{F-eqn n+1}
\B{F}(s_\alpha^{n+1},\theta^{n+1}; \Bt^n,\B{n}^n) &=&
S_b \left(D_{\Delta \alpha}s_\alpha^{n+1}\Bt^{n}
+(s_\alpha^{n+1}-1)D_{\Delta \alpha}\theta^{n+1} \B{n}^n\right)
\end{eqnarray}

It is important to note that the above discretization is not fully
implicit. In fact, both the spreading and interpolation operators
are evaluated at the interface $\B{X}^n$ from the previous time
step. Moreover, when solve the $s_\al^{n+1}$ and $\B{u}^{n+1}$,
in \myref{disc unsteady Stokes eq} - \myref{stable disc sa eq}, we use
$\theta^n$ instead of $\theta^{n+1}$ to evaluate the force density.
 This makes our semi-implicit
discretization linear with respect to the implicit solution
variables, $\B{u}^{n+1}$, $\theta^{n+1}$, and $s_\alpha^{n+1}$. The
above semi-implicit discretization essentially decouples the
stiffness induced by the elastic force from the fluid equations.
This enables us to remove the stiffness of the Immersed Boundary
method effectively by applying the Small Scale Decomposition and
arclength/tangent angle formulation as was done in \cite{HLS94}.

In the following, we will prove that this semi-implicit
discretization is unconditionally stable in the energy norm.

By using a discrete summation by parts, we can show that
\begin{equation}
< f,D_{\Delta \alpha}g>_{\Gamma_h} = - < D_{\Delta
\alpha}f,g>_{\Gamma_h} , \quad < {\bf u},\nabla_h g>_{\Omega_h} = -
< \nabla_h \cdot{\bf u}, g>_{\Omega_h} . \label{sum by part}
\end{equation}

First, we define the total energy of the physical system. The total energy
includes the kinetic energy $K$ and the potential energy $P$, which are
defined below:
\begin{eqnarray}
K&=&\frac{1}{2}\rho<\B{u},\B{u}>_{\Omega_h}=\frac{\rho}{2}\sum_{i,j=1}^{N}\B{u}_{ij}\cdot \B{u}_{ij}h^2,\\
P&=&\frac{1}{2}S_b<s_\alpha-1,s_\alpha-1>_{\Gamma_h}=\frac{S_b}{2}\sum_{j=1}^{N_b}(s_{\alpha,j}-1)^2 \Delta \alpha .
\end{eqnarray}
The total energy is then defined as
\begin{eqnarray}
E=K+P.
\end{eqnarray}

Below we will prove the unconditional stability of our semi-implicit
discretization. To simplify the presentation, we still denote the discrete spectral
derivative $D_{\Delta \alpha}g$ of a function $g$ as $g_\alpha$.

Taking the discrete inner product defined by
\myref{inner product omega} of \myref{disc unsteady Stokes eq} with
$\B{u}^{n+1}+\B{u}^{n}$ and using \myref{sum by part}, we obtain
\begin{eqnarray}
\label{kinetic}
2(K^{n+1}-K^n)&=&\rho<\B{u}^{n+1}+\B{u}^{n},\B{u}^{n+1}-\B{u}^{n}>_{\Omega_h}\nonumber\\
&=&\rho<-\B{u}^{n+1}+\B{u}^{n},\B{u}^{n+1}-\B{u}^{n}>_{\Omega_h}+2\rho<\B{u}^{n+1},\B{u}^{n+1}-\B{u}^{n}>_{\Omega_h}\nonumber\\
&=&- \rho<\B{u}^{n+1}-\B{u}^{n},\B{u}^{n+1}-\B{u}^{n}>_{\Omega_h}\nonumber\\
&&+2\Delta t(<\B{u}^{n+1},-\nabla_h p+\mu \nabla_h^2 \B{u}^{n+1}+L_{h,n}\left(\B{F}(s_\alpha^{n+1},\theta^{n})\right)>_{\Omega_h})\nonumber\\
&=&-\rho<\B{u}^{n+1}-\B{u}^{n},\B{u}^{n+1}-\B{u}^{n}>_{\Omega_h}-2\Delta t<\B{u}^{n+1},\nabla_h p>_{\Omega_h}\nonumber\\
&&+2\Delta t<\B{u}^{n+1},\mu \nabla_h^2 \B{u}^{n+1}>_{\Omega_h}+2\Delta t<\B{u}^{n+1},L_{h,n}\left(\B{F}(s_\alpha^{n+1},\theta^{n})\right)>_{\Omega_h}\nonumber\\
&=&-\rho<\B{u}^{n+1}-\B{u}^{n},\B{u}^{n+1}-\B{u}^{n}>_{\Omega_h}-2\Delta t<\nabla_h\cdot\B{u}^{n+1}, p>_{\Omega_h}\nonumber\\
&&-2\mu\Delta t<\nabla_h\B{u}^{n+1}, \nabla_h \B{u}^{n+1}>_{\Omega_h}\nonumber\\
&&+2\Delta
t<L^*_{h,n}\left(\B{u}^{n+1}\right),\B{F}(s_\alpha^{n+1},\theta^{n})>_{\Gamma_h}.
\end{eqnarray}
The second term on the right hand side of \myref{kinetic} is zero because
the discrete velocity field is divergence free, i.e.
$\nabla_h\cdot\B{u}^{n+1}=0$. The fourth term can be rewritten as
\begin{eqnarray}
\label{4th term}
&&<L^*_{h,n}\left(\B{u}^{n+1}\right),\B{F}(s_\alpha^{n+1},\theta^{n})>_{\Gamma_h}\nonumber\\
&=&<V^{n+1}\Bt^{n}+U^{n+1}\B{n}^n,S_b \left(s^{n+1}_{\al,\al}\Bt^n+(s^{n+1}_\al-1)\theta_\al^{n}\B{n}^n\right)>_{\Gamma_h}\nonumber\\
&=&S_b\left(<V^{n+1},s_{\alpha,\al}^{n+1}>_{\Gamma_h}+<U^{n+1},(s_\alpha^{n+1}-1)\theta^{n}_\al>_{\Gamma_h}\right).
\end{eqnarray}
Combining \myref{kinetic} and \myref{4th term}, we can get
\begin{eqnarray}
\label{final kinetic} 2(K^{n+1}-K^n)
&=&-\rho<\B{u}^{n+1}-\B{u}^{n},\B{u}^{n+1}-\B{u}^{n}>_{\Omega_h}-2\mu\Delta
t<\nabla_h\B{u}^{n+1}, \nabla_h \B{u}^{n+1}>_{\Omega_h}\nonumber
\\
&&+2S_b\Delta
t(<V^{n+1},s_{\alpha,\al}^{n+1}>_{\Gamma_h}+<U^{n+1},(s_\alpha^{n+1}-1)\theta_\al^{n}>_{\Gamma_h}).
\end{eqnarray}
Similarly, by taking the discrete inner product defined by
\myref{inner product gamma} of \myref{stable disc sa eq} with
$s_\al^{n+1}+s_\al^{n}-2$ and using \myref{sum by part}, we get
\begin{eqnarray}
\label{potential}
2(P^{n+1}-P^n)&=&S_b<s_\alpha^{n+1}+s_\alpha^{n}-2,s_\alpha^{n+1}-s_\alpha^{n}>_{\Gamma_h}\nonumber\\
&=&S_b<-s_\alpha^{n+1}+s_\alpha^{n},s_\alpha^{n+1}-s_\alpha^{n}>_{\Gamma_h}+2S_b<s_\alpha^{n+1}-1,s_\alpha^{n+1}-s_\alpha^{n}>_{\Gamma_h}\nonumber\\
&=&-S_b <s_\alpha^{n+1}-s_\alpha^{n},s_\alpha^{n+1}-s_\alpha^{n}>_{\Gamma_h}\nonumber\\
&&+2 S_b\Delta t<s_\alpha^{n+1}-1,V_\al^{n+1}-\theta_\al^{n}U^{n+1}>_{\Gamma_h}\nonumber\\
&=&-S_b <s_\alpha^{n+1}-s_\alpha^{n},s_\alpha^{n+1}-s_\alpha^{n}>_{\Gamma_h}\nonumber\\
&&+2 S_b\Delta
t(-<s_{\alpha,\al}^{n+1},V^{n+1}>_{\Gamma_h}-<s_\alpha^{n+1}-1,\theta_\al^{n}U^{n+1}>_{\Gamma_h}).
\end{eqnarray}
Adding \myref{final kinetic} to \myref{potential}, we have
\begin{eqnarray}
E^{n+1}-E^n&=&-\frac{1}{2}\rho<\B{u}^{n+1}-\B{u}^{n},\B{u}^{n+1}-\B{u}^{n}>_{\Omega_h}-\mu\Delta t <\nabla_h\B{u}^{n+1},\nabla_h\B{u}^{n+1}>_{\Omega_h}\nonumber\\
&&-\frac{1}{2}S_b<s_\alpha^{n+1}-s_\alpha^{n},s_\alpha^{n+1}-s_\alpha^{n}>_{\Gamma_h} \nonumber \\
&\leq & 0 .
\end{eqnarray}
This proves that our semi-implicit discretization is unconditionally stable
in the sense that the total energy is non-increasing.

\vspace{0.1in} \noindent {\bf Remark 1.} In our proof presented
above, we have used two important properties of our semi-implicit
discretization. The first property is that the
 discrete spreading and interpolation operators are adjoint. The second
property is that the velocity field satisfies the discrete
divergence free condition. It is clear from the above proof that as
long as these two properties are satisfied by our spatial
discretization, the corresponding semi-implicit discretization
introduced in the previous subsection is unconditionally stable.

\vspace{0.1in} \noindent {\bf Remark 2.} We remark that the proof
above is similar in spirit to that of a semi-linear discretization
obtained by Newrenn et al in \cite{NFGK07}. There is some minor
difference between the unconditionally stable semi-implicit
discretization obtained by Newren et al and our unconditionally
stable semi-implicit discretization. In the problem considered
by Newren et al., the force is a linear function of the interface.
On the other hand, in the problem we consider, the force is a
nonlinear function of the interface (the rest length of the
boundary is not zero). By using the $s_\al-\theta$ formulation,
the force is a linear function of $s_\al$. By treating $\theta$
explicitly, we obtain a semi-implicit discretization that is
linear with respect to $s_\al$. Due to the decoupling
between $s_\al$ and $\theta$, we need to solve two
$N_b\times N_b$ linear systems instead of one $2N_b\times
2N_b$ linear system in the semi-implicit discretization obtained
by Newren et al.

\vspace{0.1in} \noindent {\bf Remark 3.} For the steady Stokes flow,
we can also prove the following semi-implicit discretization is
unconditionally stable:

\vspace{0.1in}
{\bf Step 1}:
\begin{eqnarray}
\label{dis steady Stokes}
0&=&-\nabla_h p^{n+1}+\mu \nabla_h^2 \B{u}^{n+1}+L_{h,n}(\B{F}(s_\alpha^{n+1},\theta^{n})),\\
\nabla_h^2 p^{n+1}&=&\nabla_h\cdot L_{h,n}(\B{F}(s_\alpha^{n+1},\theta^{n}; \Bt^n,\B{n}^n))), \\
V^{n+1}&=&L^*_{h,n}(\B{u}^{n+1}(\B{x}))\cdot \Bt^{n},\\
U^{n+1}&=&L^*_{h,n}(\B{u}^{n+1}(\B{x}))\cdot \B{n}^{n},\\
\label{eq of steady s} \frac{s_\alpha^{n+1}-s_\alpha^{n}}{\Delta
t}&=&
D_{\Delta \alpha} V^{n+1}- D_{\Delta \alpha}\theta^{n}U^{n+1},\\
\label{eq of steady theta} \frac{\theta^{n+1}-\theta^{n}}{\Delta
t}&=&\frac{1}{s_\alpha^{n+1}} \left(D_{\Delta \alpha}
U^{n+1}+D_{\Delta \alpha}\theta^{n+1}V^{n+1}\right) .
\end{eqnarray}

\vspace{0.1in}
{\bf Step 2}:
\begin{eqnarray}
\label{disc vel theta steady}
0&=&
-\nabla_h \overline{p}^{\;n+1}+\mu \nabla_h^2 \overline{\B{u}}^{n+1}
+L_{h,n}\left(\B{F}(s_\alpha^{n+1},\theta^{n+1}; \Bt^n,\B{n}^n)\right), \\
\label{disc p theta steday}
\nabla_h^2 \;\overline{p}^{\;n+1}&=&\nabla_h\cdot
L_{h,n}\left(\B{F}(s_\alpha^{n+1},\theta^{n+1}; \Bt^n,\B{n}^n)\right)
, \\
\overline{V}^{\;n+1}&=&L^*_{h,n}(\overline{\B{u}}^{\;n+1})\cdot \Bt^{n} \\
\overline{U}^{\;n+1}&=& L^*_{h,n}(\overline{\B{u}}^{\;n+1})\cdot \B{n}^{n},\\
\label{stable disc theta eq steady}
\frac{\theta^{n+1}-\theta^{n}}{\Delta t}&=&\frac{1}{s_\alpha^{n+1}}
\left(D_{\Delta \alpha}\overline{U}^{\;n+1}+D_{\Delta \alpha}\theta^{n}\overline{V}^{\;n+1}\right).
\end{eqnarray}
In this case the total energy is just the potential energy
\begin{eqnarray}
E=\frac{S_b}{2}<s_\al-1,s_\al-1>_{\Gamma_h}=\frac{S_b}{2}\sum_{j=1}^{N_b}(s_{\al,j}-1)^2
\Delta \alpha.
\end{eqnarray}
As in the case of the unsteady Stokes flow, as long as the velocity
field satisfies the discrete divergence free condition and the
discrete spreading and interpolation operators are adjoint, we can
prove that above semi-implicit discretization is unconditionally
stable in the sense of total energy is non-increasing.

\subsection{Small Scale Decomposition}

In order to apply the Small Scale Decomposition to our unconditionally
stable time discretization, we would like to solve for the velocity
field at time $t^{n+1}$ from the space-continuous version of \myref{disc unsteady Stokes eq} and
\myref{disc p} using an integral representation:
\begin{eqnarray}
\B{u}^{n+1}(\B{x})&=&\left(1-\frac{\mu \Delta t}{\rho}\nabla^2\right)^{-1}\left(\B{u}^{n}+\frac{\Delta t}{\rho}(1-\nabla(\nabla^2)^{-1}\nabla\cdot) L_{n}(\B{F}(s_\alpha^{n+1},\theta^{n}))\right)\nonumber\\
&=&\left(1-\frac{\mu \Delta t}{\rho}\nabla^2\right)^{-1}\left(\B{u}^{n}+\frac{\Delta t}{\rho} L_{n}(\B{F}(s_\alpha^{n+1},\theta^{n}))\right)\nonumber\\
&&-\frac{\Delta t}{\rho}\left(1-\frac{\mu \Delta t}{\rho}\nabla^2\right)^{-1}(\nabla^2)^{-1}\left(\nabla\nabla\cdot L_{n}(\B{F}(s_\alpha^{n+1},\theta^{n}))\right)\nonumber .
\end{eqnarray}
To solve for the velocity field at $t^{n+1}$, we need to use the following
free space fundamental solutions in two space dimensions which are defined
as follows:
\begin{eqnarray}
\left(1-\frac{\mu \Delta t}{\rho}\nabla^2\right)E_1=\delta(\B{x}-\B{x}'),\\
\nabla^2\left(1-\frac{\mu \Delta t}{\rho}\nabla^2\right)E_2=\delta(\B{x}-\B{x}').
\end{eqnarray}
These two fundamental solutions can be expressed in terms of the
modified Bessel function of the second kind \cite{AS72}:
\begin{eqnarray}
E_1&=&\frac{\lambda^2}{2\pi} K_0(\lambda|\B{x}-\B{x}'|),\\
E_2&=&\frac{1}{2\pi}\left( K_0(\lambda|\B{x}-\B{x}'|+
\ln(|\B{x}-\B{x}'|)\right),
\end{eqnarray}
where $\D\lambda^2=\frac{\rho}{\mu \Delta t}$ and $K_0$ is
a modified Bessel function of the second kind.
By integrating by part, we can further express the velocity
$\B{u}^{n+1}$ as
\begin{eqnarray}
\label{unsteady vel}
\B{u}^{n+1}(\B{x})&=&\frac{1}{2\pi}\int_\Omega \lambda^2 K_0(\lambda|\B{x}-\B{x}'|)\B{u}^n(\B{x}')d\B{x}'\nonumber\\
&&+\frac{1}{2\pi}\frac{\Delta t}{\rho} \int_{\Gamma^n} \lambda^2 K_0(\lambda|\B{x}-\B{X}^n(\alpha')|)\B{F}(s_\alpha^{n+1},\theta^{n})d\alpha'\nonumber\\
&&-\frac{1}{2\pi}\frac{\Delta t}{\rho}\int_\Omega \nabla_\B{x}\nabla_\B{x}(K_0(\lambda|\B{x}-\B{x}'|)+\ln(|\B{x}-\B{x}'|))
\cdot L^n(\B{F}(s_\alpha^{n+1},\theta^{n}))d\B{x}'\nonumber\\
&=&\frac{1}{2\pi}\int_\Omega \lambda^2 K_0(\lambda|\B{x}-\B{x}'|)\B{u}^n(\B{x}')d\B{x}'\nonumber\\
&&+\frac{1}{2\pi}\frac{\Delta t}{\rho} \int_{\Gamma^n} \lambda^2 K_0(\lambda|\B{x}-\B{X}^n(\alpha')|)\B{F}(s_\alpha^{n+1},\theta^{n})d\alpha'\nonumber\\
&&-\frac{1}{2\pi} \frac{\Delta t}{\rho}\int_\Omega G(\B{x}-\B{X}^n(\alpha'))
\cdot \B{F}(s_\alpha^{n+1},\theta^{n})d\alpha',
\end{eqnarray}
where $G$ is defined as follows:
\begin{eqnarray}
\label{unsteady kernel}
G_{ij}(\B{r})&=&\frac{\delta_{ij}}{|\B{r}|^2}-\frac{2r_ir_j}{|\B{r}|^4}+\frac{1}{2}\lambda^2(K_0(\lambda|\B{r}|)+K_2(\lambda|\B{r}|))\frac{r_ir_j}{|\B{r}|^2}\nonumber\\
&&-\lambda K_1(\lambda|\B{r}|)\left(\frac{\delta_{ij}}{|\B{r}|}-\frac{r_ir_j}{|\B{r}|^3}\right),
\end{eqnarray}
and $K_0, K_1, K_2$ are all modified Bessel functions of the second kind
\cite{AS72}.

In this subsection, we will perform a Small Scale Decomposition to
the velocity field based on the integral representation \myref{unsteady vel}.
Recall that in our semi-implicit discretization, the velocity field
at $t^{n+1}$ is evaluated on the boundary $\B{X}^n$ at $t^n$. Thus
we should perform our Small Scale Decomposition for $\B{u}^{n+1}(\B{X}^n)$.
To this end, we first write down the integral expression of
$\B{u}^{n+1}(\B{X}^n)$ as follows:
\begin{eqnarray}
\B{u}^{n+1}(\B{X}^n(\alpha))&=&\frac{1}{2\pi}\int_\Omega \lambda^2 K_0(\lambda|\B{X}^n(\alpha)-\B{x}'|)\B{u}^n(\B{x}')d\B{x}'\nonumber\\
&&+\frac{1}{2\pi}\frac{\Delta t}{\rho} \int_{\Gamma^n} \lambda^2 K_0(\lambda|\B{X}^n(\alpha)-\B{X}^n(\alpha')|)\B{F}(s_\alpha^{n+1},\theta^{n})d\alpha'\nonumber\\
&&-\frac{1}{2\pi} \frac{\Delta t}{\rho}\int_\Omega G(\B{X}^n(\alpha)-\B{X}^n(\alpha'))
\cdot \B{F}(s_\alpha^{n+1},\theta^{n})d\alpha'.
\end{eqnarray}
To perform the Small Scale Decomposition to the above velocity integral,
we would like to decompose the singular velocity kernel as the sum of a
linear singular operator of convolution type and a remainder operator
which is regular. Using the Taylor expansion for $\alpha'$ around
$\alpha$, we get the following decomposition:
\begin{eqnarray}
\label{leading of unsteady V}
&&V^{n+1}(\alpha)=
\B{u}^{n+1}(\B{X}^n(\alpha))\cdot \Bt^n(\alpha)\nonumber\\
&\sim& \frac{S_b\Delta t}{2\pi\rho} \int_{\Gamma^n} \lambda^2 K_0(\lambda s^n_\alpha|\alpha-\alpha'|)s^{n+1}_{\alpha,\alpha'}d\alpha'-\nonumber\\
&&\frac{S_b\Delta t}{2\pi\rho}\int_{\Gamma^n} \left(\frac{1}{2}\lambda^2(K_0(\lambda s^n_\alpha|\alpha-\alpha'|)+K_2(\lambda s^n_\alpha|\alpha-\alpha'|))-\frac{1}{\left(s_\alpha^{n}\right)^2(\alpha-\alpha')^2}\right)\nonumber\\
&&\quad \quad \quad \quad \quad s^{n+1}_{\alpha,\alpha'}d\alpha',
\end{eqnarray}
where $s^n_\alpha|$ inside $K_0(\lambda s^n_\alpha|\alpha-\alpha'|)$ is
evaluated at $\al$. Notice that \cite{AS72}
\begin{eqnarray}
\frac{d^2}{d\al'^2}\left(\frac{1}{\left(s^n_\alpha(\al)\right)^2}K_0(\lambda s^n_\alpha|\alpha-\alpha'|)\right)=\frac{1}{2}\lambda^2\left(K_0(\lambda s^n_\alpha|\alpha-\alpha'|)+K_2(\lambda s^n_\alpha|\alpha-\alpha'|)\right).
\end{eqnarray}
Integrating the right hand side of \myref{leading of unsteady V} by parts
twice, we get
\begin{eqnarray}
\label{leading of unsteady V final}
V^{n+1}(\alpha)&\sim&\frac{1}{2\pi\rho}S_b\Delta t \int_{\Gamma^n} \lambda^2 K_0(\lambda s^n_\alpha|\alpha-\alpha'|)s^{n+1}_{\alpha,\alpha'}d\alpha'-\nonumber\\
&&\frac{S_b\Delta t}{2\pi\rho \left(s^n_\alpha\right)^2} \int_{\Gamma^n}
\left(K_0(\lambda
s^n_\alpha|\alpha-\alpha'|)-\ln(\alpha-\alpha')\right)s^{n+1}_{\alpha,\alpha'\alpha'\alpha'}d\alpha'.
\end{eqnarray}
Similarly, we can obtain the leading order contribution of $\overline{U}^{\;n+1}$ as
follows:
\begin{eqnarray}
\label{leading of unsteady U}
&&\overline{U}^{\;n+1}(\alpha)=\overline{\B{u}}^{\;n+1}(\B{X}^n(\alpha))\cdot \B{n}^n(\alpha)\nonumber\\
&&\sim\frac{S_b\Delta t}{2\pi\rho \left(s^n_\alpha\right)^2}
\int_{\Gamma^n} \left(K_0(\lambda
s^n_\alpha|\alpha-\alpha'|)-\ln(\alpha-\alpha')\right)\left((s^{n+1}_{\alpha}-1)\theta^{n+1}_{\alpha'}\right)_{\alpha'\alpha'}d\alpha'.
\end{eqnarray}
Using this decomposition, we obtain the following scheme:
\begin{eqnarray}
\label{semi implicit sa} \frac{s_\alpha^{n+1}-s_\alpha^{n}}{\Delta
t}&=&T(s_\al^{n+1})+\left( D_{\Delta \alpha}V^{*,n+1}-D_{\Delta
\alpha}\theta^{n}U^{*,n+1}-
T(s_\al^n)\right), \\
\rho\frac{\B{u}^{n+1}-\B{u}^n}{\Delta t}&=&-\nabla_h p^{n+1}
+\mu \nabla_h^2 \B{u}^{n+1}+L_{h,n}(\B{F}(s_\alpha^{n+1},\theta^{n})),\\
\nabla_h^2 p^{n+1}&=&\nabla_h\cdot L_{h,n}(\B{F}(s_\alpha^{n+1},\theta^{n})), \\
V^{n+1}&=&L^*_{h,n}(\B{u}^{n+1})\cdot \Bt^{n},\\
U^{n+1}&=&L^*_{h,n}(\B{u}^{n+1})\cdot \B{n}^{n},\\
\frac{\theta^{n+1}-\theta^{n}}{\Delta t}&=&\frac{S(\theta^{n+1})}{s_\al^{n+1}}
+\left(\frac{1}{s_\alpha^{n+1}}\left(D_{\Delta \alpha}U^{n+1}
+ D_{\Delta \alpha}\theta^{n}V^{n+1}\right)-S(\theta^n)\right),
\end{eqnarray}
where
\begin{eqnarray*}
T(s_\al^{n+1})&=&\left(\frac{1}{2\pi\rho}S_b\Delta t \int_{\Gamma^n} \lambda^2 K_0(\lambda s^n_\alpha|\alpha-\alpha'|)s^{n+1}_{\alpha,\alpha'}d\alpha'\nonumber\right)_\al\\
&&-\left(\frac{S_b\Delta t}{2\pi \rho\left(s^n_\alpha\right)^2} \int_{\Gamma^n} \left(K_0(\lambda s^n_\alpha|\alpha-\alpha'|)-\ln(\alpha-\alpha')\right)s^{n+1}_{\alpha,\alpha'\alpha'\alpha'}d\alpha'\right)_\al ,\\
S(\theta^{n+1})&=&\left(\frac{S_b\Delta t}{2\pi\rho
\left(s^n_\alpha\right)^2} \int_{\Gamma^n} \left(K_0(\lambda
s^n_\alpha|\alpha-\alpha'|)-\ln(\alpha-\alpha')\right)\left((s^{n+1}_{\alpha}-1)\theta_{\alpha'}\right)_{\alpha'\alpha'}d\alpha'\right)_\al,
\end{eqnarray*}
and $\B{u}^{*,n+1}$ is the velocity at $t^{n+1}$ which is calculated
explicitly
\begin{eqnarray}
\label{velocity ustar}
\rho\frac{\B{u}^{*,n+1}-\B{u}^n}{\Delta t}&=&-\nabla_h p^{*,n+1}+\mu \nabla_h^2 \B{u}^{*,n+1}+L_{h,n}(\B{F}(s_\alpha^{n},\theta^{n}))\\
\label{pstar}
\nabla_h^2 p^{*,n+1}&=&\nabla_h\cdot L_{h,n}(\B{F}(s_\alpha^{n},\theta^{n})) \\
V^{*,n+1}&=&L^*_{h,n}(\B{u}^{*,n+1})\cdot
\Bt^{n}\\
\label{Ustar}
U^{*,n+1}&=&L^*_{h,n}(\B{u}^{*,n+1})\cdot \B{n}^{n}.
\end{eqnarray}
The derivation of the above semi-implicit scheme is given in Appendix $
\B{B}$.

However, the expressions of $T$ and $S$ are still too complicated and
need to be further simplified.
The leading order linear operator, which contains $K_0(\lambda
s^n_\alpha(\al)|\alpha-\alpha'|)$, is not a convolution operator.
Thus, it does not have a simple kernel under the Fourier transform
as the Hilbert operator in the case of the steady Stokes flow.
To further simplify the kernel, we approximate
$s^n_\al(\al)$ by $\D\min_\al s^n_\al(\al)$. With this approximation,
the corresponding leading order operator is a convolution operator
and can be diagonalized under the Fourier transform.
Denote $\beta=\lambda\D\min_\al s^n_\al(\al)$. In Appendix $\B{C}$,
we will show that
\begin{eqnarray}
\label{fft of K0}
\mathcal{F}\left(\frac{1}{\pi}\int_{-\infty}^{+\infty}K_0(\beta
|\al-\al'|)f(\al')d\al'\right)
=\frac{\widehat{f}(k)}{\sqrt{\beta^2+k^2}} .
\end{eqnarray}
Using \myref{fft of K0} and replacing $s^n_\al(\al)$ by $\D\min_\al
s^n_\al(\al)$, we can simplify the leading order term
$T(s_\al^{n+1})$ and $S(\theta^{n+1})$ under the Fourier transform:
\begin{eqnarray}
\label{fft of T}
\widehat{T}(s_\al^{n+1})&\sim&-\frac{S_b\Delta t}{2\rho\left(\D\min_\al s^n_\al\right)^2}\left(\frac{\left(\lambda\D\min_\al s^n_\al\right)^2k^2+k^4}{\sqrt{\left(\lambda\D\min_\al s_\al^n\right)^2+k^2}}-|k|^3\right)\hat{s}^{n+1}_\al,\\
\label{fft of S}
\widehat{S}(\theta^{n+1})&\sim&-\frac{S_b\Delta t\D\max_\al\left(s^{n+1}_\al-1\right)}{2\rho\left(\D\min_\al s^n_\al\right)^2}\left(|k|^3-\frac{k^4}{\sqrt{\left(\lambda\D\min_\al s_\al^n\right)^2+k^2}}\right)\hat{\theta}^{n+1}.
\end{eqnarray}
When $\mu\gg 1$, we have $\D\lambda=\frac{1}{\sqrt{\mu\Delta t}}\ll 1$.
By Taylor expanding \myref{fft of T} and \myref{fft of S} with
respect to $\lambda$ and keeping only the first order term,
we obtain the leading order term as follows:
\begin{eqnarray}
\widehat{T}(s_\al^{n+1})&\sim&-\frac{S_b}{4\mu}|k|\hat{s}^{n+1}_\al,\\
\widehat{S}(\theta^{n+1})&\sim&-\frac{S_b}{4\mu}\D\max_\al\left(s^{n+1}_\al-1\right)|k|\hat{\theta}^{n+1},
\end{eqnarray}
which is the same as the steady Stoke flow.
This is also consistent with one's physical intuition. When the
viscosity is very large, the flow changes very slowly. The inertial
term can be neglected.

When $\mu\ll 1$, then $\D\lambda=\frac{1}{\sqrt{\mu\Delta t}}\gg 1$, the asymptotic expansion is
\begin{eqnarray}
\widehat{T}(s_\al^{n+1})&\sim&-\frac{S_b\sqrt{\Delta t}}{2\left(\D\min_\al s^n_\al\right)\sqrt{\rho\mu}}k^2\hat{s}^{n+1}_\al,\\
\widehat{S}(\theta^{n+1})&\sim&-\frac{S_b\,\Delta t\D\max_\al\left(s^{n+1}_\al-1\right)}{2\rho\left(\D\min_\al s^n_\al\right)^2}k^3\hat{\theta}^{n+1}.
\end{eqnarray}

From the asymptotic expansion above, we can see that our small scale decomposition is also consistent with the linearized stability analysis
which Stockie and Wetton got in \cite{SW95}. Using the leading order term above, we can get
the leading order term of the eigenvalue same with the result in \cite{SW95}.

We can also obtain the corresponding
stability constraint for the explicit scheme near the equilibrium:
\begin{eqnarray}
\Delta t < C(S_b,\mu)h^\beta,
\end{eqnarray}
where $1\leq \beta \leq 3/2$. The value of $\beta$ depends on
$\mu$. If $\mu\ll 1$, then we have $\beta \approx 3/2$.
On the other hand, if $\mu\gg 1$, we have $\beta \approx 1$.

\subsection{The numerical scheme}

Based on the small scale decomposition we developed in the last
subsection, we can now describe our semi-implicit numerical scheme.
Combining the time discretization \myref{disc unsteady Stokes eq}
-\myref{stable disc theta eq} with the decomposition
\myref{leading of unsteady V}-\myref{leading of unsteady U} and
using the approximation \myref{fft of T}-\myref{fft of S}, we
obtain the following semi-implicit numerical scheme:

\vspace{0.1in}
{\bf Step 1}: Update of ${\bf u}^{n+1}$ and $s_\alpha^{n+1}$.
\begin{eqnarray}
\frac{s_\alpha^{n+1}-s_\alpha^{n}}{\Delta t}&=&T(s_\al^{n+1})+\left(
D_{\Delta \alpha}V^{*,n+1}- D_{\Delta \alpha}\theta^{n}U^{*,n+1}-T(s_\al^n)\right),\\
\rho\frac{\B{u}^{n+1}-\B{u}^n}{\Delta t}&=&-\nabla_h
p^{n+1}+\mu \nabla_h^2 \B{u}^{n+1}+L_{h,n}(\B{F}(s_\alpha^{n+1},\theta^{n})),\\
\nabla_h^2 p^{n+1}&=&\nabla_h\cdot L_{h,n}(\B{F}(s_\alpha^{n+1},\theta^{n})),
\end{eqnarray}
where
\begin{eqnarray}
\label{old unsteady leading s}
\widehat{T}(s_\al^{n+1})&=&-\frac{S_b\Delta t}{2\rho\left(\D\min_\al s^n_\al\right)^2}\left(\frac{\left(\lambda\D\min_\al s_\al^n\right)^2k^2+k^4}{\sqrt{\left(\lambda\D\min_\al s_\al^n\right)^2+k^2}}-|k|^3\right)\hat{s}^{n+1}_\al,\\
\end{eqnarray}
and $\B{u}^{*,n+1}$ is the intermediate velocity at $t^{n+1}$ which is
calculated by solving the unsteady Stokes equations implicitly
while evaluating the elastic force explicitly:
\begin{eqnarray}
\rho\frac{\B{u}^{*,n+1}-\B{u}^n}{\Delta t}&=&-\nabla_h p^{*,n+1}+\mu
\nabla_h^2 \B{u}^{*,n+1}+L_{h,n}(\B{F}(s_\alpha^{n},\theta^{n})),\\
\nabla_h^2 p^{*,n+1}&=&\nabla_h\cdot L_{h,n}(\B{F}(s_\alpha^{n},\theta^{n})), \\
V^{*,n+1}&=&L^*_{h,n}(\B{u}^{*,n+1})\cdot
\Bt^{n},\\
U^{*,n+1}&=&L^*_{h,n}(\B{u}^{*,n+1})\cdot \B{n}^{n}.
\end{eqnarray}

\vspace{0.1in}
{\bf Step 2}: Update of $\theta^{n+1}$.
Once we have updated ${\bf u}$, $p$, and $s_\alpha$ at $t^{n+1}$, we
update $\theta^{n+1}$ using the following semi-implicit scheme:
\begin{eqnarray}
\frac{\theta^{n+1}-\theta^{n}}{\Delta t}&=&\frac{S(\theta^{n+1})}{\min_\al s_\al^{n+1}}+\left(\frac{1}{s_\alpha^{n+1}}\left(D_{\Delta \alpha}U^{n+1}+D_{\Delta \alpha}\theta^{n}V^{n+1}\right)-S(\theta^n)\right),
\end{eqnarray}
where
\begin{eqnarray}
V^{n+1}&=&L^*_{h,n}(\B{u}^{n+1})\cdot \Bt^{n},\\
U^{n+1}&=&L^*_{h,n}(\B{u}^{n+1})\cdot \B{n}^{n},\\
\label{old unsteady leading t}
\widehat{S}(\theta^{n+1})&=&-\frac{S_b\Delta t\D\max_\al\left(s^n_\al-1\right)}{2\rho\left(\D\min_\al s^n_\al\right)^2}\left(|k|^3-\frac{k^4}{\sqrt{\left(\lambda\D\min_\al s_\al^n\right)^2+k^2}}\right)\hat{\theta}^{n+1}.
\end{eqnarray}

This is our semi-implicit scheme for the unsteady Stokes
flow. The spectral discretization in space has the advantage
of being high order accurate and the leading order operator has
a simple kernel under the Fourier transform. As it is, the time
discretization is only first order. Based on the first order
semi-implicit scheme that we develop in this subsection, we will
develop a second order semi-implicit scheme in the next subsection.

A near equilibrium stability analysis shows that the stability
constraint of this semi-implicit scheme is of the form $\Delta t<
C(S_b,\mu)$, which is independent of the wave number, but still
dependent on $S_b$ and $\mu$. This is due to the fact that the Small
Scale Decomposition does not capture the low frequency components of
the solution accurately. The low frequency components of the
solution can affect the stability of the time discretization in two
ways. The first one is through the small scale decomposition, which
only captures the leading order contribution of the solution at high
wave numbers. The second one comes from the second term of the right
hand side of the dynamic equations for $s_\al$ and $\theta$. As in
the case of the steady Stokes flow, we can include the leading order
contribution from the second term in our leading order term and
treat them implicitly. This treatment would significantly improve
the stability property especially when the elastic coefficient is
large or the viscosity is small. This improved stability is
at the expense of solving a linear system for the implicit
solution at each time step. We call this semi-implicit discretization
as the semi-implicit method of the second kind. More discussions
on the semi-implicit method of the second kind can be found in
Appendix $\B{A}$.

\vspace{0.1in} 
\noindent 
{\bf Remark 4.} 
The leading order term we derive above is calculated analytically using the 
space-continuous formulation with an unsmoothed Dirac delta function. As 
Stockie and Wetton pointed out in \cite{SW99}, this analysis over-predicts 
the stiffness of the Immersed Boundary method in a practical computation. 
If we use the leading order approximation directly, the semi-implicit scheme 
with the leading order terms derived above tends to over-dissipate the 
solution. To alleviate this effect in the practical implementation, we 
rescale the leading order term by a coefficient which is calculated at 
the first time step in the following way: 
\begin{eqnarray}
C_V=\frac{\max_\al V_\al^{1,*}}{\max_\al T\left(s_\al^0\right)},\nonumber\\
C_U=\frac{\max_\al U^{1}}{\max_\al S_U\left(\theta^0\right)}\nonumber,
\end{eqnarray}
where $S_U(\theta^{0})$ is the leading order term of $\overline{U}^{\;1}$,
which can be computed from $S(\theta^{0})$ via the Fourier transform.
The leading order term we use in a practical computation is 
actually $C_V T(s_\al^{n+1})$ and $C_U S(\theta^{n+1})$.
 
\subsection{A second order semi-implicit scheme}

Based on the first order semi-implicit scheme we have developed in the
previous subsection, we will derive the corresponding second order
semi-implicit scheme in this subsection.

First, we need to use a robust implicit second order temporal
discretization. To simplify the presentation, we will only describe
the semi-discrete algorithm. The space discretization is done in the
same way as before. The second order temporal discretization we use
consists of two steps. In the first step, we take a fractional
time step from $t^n$ to $t^{n+\frac{1}{2}}$. It is same with the first order
semi-implicit discretization \myref{disc unsteady Stokes eq}-\myref{stable disc theta eq}
, except the timestep is $\frac{\Delta t}{2}$.

In the second step, we integrate the unsteady Stokes equations
from $t^n$ to $t^{n+1}$ based on the midpoint and the trapezoidal rules:

\vspace{0.1in}
{\bf Step 1}: Update of ${\bf u}^{n+1}$ and $s_\alpha^{n+1}$.
\begin{eqnarray}
\label{main step of 2nd order scheme}
\rho\frac{\B{u}^{n+1}-\B{u}^n}{\Delta t}&=&-\nabla \bar{p}+
\mu \nabla^2 \bar{\B{u}}+L_{n+\frac{1}{2}}
\left(\B{F}\left(\bar{s}_\alpha,\theta^{n+\frac{1}{2}}; \Bt^{n+\frac{1}{2}},\B{n}^{n+\frac{1}{2}}\right)\right),\\
\nabla^2 \bar{p}&=&\nabla \cdot L_{n+\frac{1}{2}}\left(\B{F}
\left(\bar{s}_\alpha,\theta^{n+\frac{1}{2}}; \Bt^{n+\frac{1}{2}},\B{n}^{n+\frac{1}{2}}\right)\right) ,\\
\bar{V}&=&L^*_{n+\frac{1}{2}}\left(\bar{\B{u}}\right)\cdot
\Bt^{n+\frac{1}{2}},\\
\bar{U}&=&L^*_{n+\frac{1}{2}}\left(\bar{\B{u}}\right)\cdot
\B{n}^{n+\frac{1}{2}},\\
\frac{s_\alpha^{n+1}-s_\alpha^{n}}{\Delta t/2}&=&
\bar{V}_\alpha-\theta_\alpha^{n+\frac{1}{2}}\bar{U},
\end{eqnarray}
where $\Bt^{n+\frac{1}{2}} = (\cos(\theta^{n+\frac{1}{2}}),\sin(\theta^{n+\frac{1}{2}}))$,
$\B{n}^{n+\frac{1}{2}} = (-\sin(\theta^{n+\frac{1}{2}}), \cos(\theta^{n+\frac{1}{2}}))$,
 $L_{n+\frac{1}{2}}=L(\B{X}^n+\frac{1}{2}), L^*_{n+\frac{1}{2}}=L^*(\B{X}^n+\frac{1}{2})$
and
\begin{eqnarray}
\label{F-eqn n+1/2}
\B{F}(\bar{s}_\alpha,\theta^{n+\frac{1}{2}}; \Bt^{n+\frac{1}{2}},\B{n}^{n+\frac{1}{2}}) &=&
S_b \left(D_{\Delta \alpha}\bar{s}_\alpha\Bt^{n+\frac{1}{2}}
+(\bar{s}_\alpha-1)D_{\Delta \alpha}\theta^{n+\frac{1}{2}} \B{n}^{n+\frac{1}{2}}\right)
\end{eqnarray}

\vspace{0.1in}
{\bf Step 2}: Update of $\theta^{n+1}$.
After we have obtained ${\bf u}^{n+1}$ and
$s_\alpha^{n+1}$, we update $\theta$ at $t^{n+1}$ using the following
semi-implicit scheme:
\begin{eqnarray}
\label{disc vel theta main}
\rho\frac{\B{\widetilde{u}}^{\;n+1}-\B{u}^{n}}{\Delta t}&=& -\nabla
\widetilde{p}^{\;n+1}+\mu \nabla^2
\left(\frac{\widetilde{\B{u}}^{\;n+1}+\B{u}^{\;n+1}}{2}\right)
+L_{n+\frac{1}{2}}\left(\B{F}(\bar{s}_\alpha,\bar{\theta}; \Bt^{n+\frac{1}{2}},\B{n}^{n+\frac{1}{2}})\right), \quad\\
\label{disc p theta main} \nabla^2
\;\widetilde{p}^{\;n+1}&=&\nabla\cdot
L_{n+\frac{1}{2}}\left(\B{F}(\bar{s}_\alpha,\bar{\theta};
\Bt^{n+\frac{1}{2}},\B{n}^{n+\frac{1}{2}})\right)
, \\
\widetilde{V}&=&L^*_{n+\frac{1}{2}}\left(\frac{\widetilde{\B{u}}^{\;n+1}+\B{u}^{\;n+1}}{2}\right)\cdot \Bt^{n+\frac{1}{2}} \\
\widetilde{U}&=& L^*_{n+\frac{1}{2}}\left(\frac{\widetilde{\B{u}}^{\;n+1}+\B{u}^{\;n+1}}{2}\right)\cdot \B{n}^{n+\frac{1}{2}},\\
\label{stable disc theta eq main}
\frac{\theta^{n+1}-\theta^{n}}{\Delta
t}&=&\frac{1}{s_\alpha^{n+\frac{1}{2}}}
\left(\widetilde{U}_\al+\theta_\al^{n+\frac{1}{2}}\widetilde{V}\right).
\end{eqnarray}
where
\begin{eqnarray}
\label{F-eqn bar} \B{F}(\bar{s}_\alpha,\bar{\theta};
\Bt^{n+\frac{1}{2}},\B{n}^{n+\frac{1}{2}}) &=& S_b
\left(\bar{s}_{\alpha,\al}\Bt^{n+\frac{1}{2}}
+(\bar{s}_\alpha-1)\bar{\theta}_\al \B{n}^{n+\frac{1}{2}}\right)
\end{eqnarray}
and
\begin{eqnarray}
\bar{\B{u}}=\frac{\B{u}^{n+1}+\B{u}^{n}}{2},\quad\bar{s}_\alpha=\frac{s_\alpha^{n+1}+s_\alpha^{n}}{2},\quad\bar{\theta}=\frac{\theta^{n+1}+\theta^{n}}{2}.
\end{eqnarray}
Here, $L_{n+\frac{1}{2}}$ and $L^*_{n+\frac{1}{2}}$ are the
spreading and the interpolation operators evaluated at $\B{X}^{n+\frac{1}{2}}$.
Using the same method of analysis, we can prove that the above second order
semi-implicit discretization is unconditionally stable in the sense
that the total energy is non-increasing.

The first step in our second order method is identical to the first order
method except that the time step is $\frac{\Delta t}{2}$ instead of
$\Delta t$. Thus we can use the first order semi-implicit scheme
introduced in last subsection to compute it directly.

In the second step, we can also apply the Small Scale Decomposition
with some modifications. After applying the Small Scale Decomposition
to the second step of the two-step method, the
second step of the semi-implicit scheme has the form

\vspace{0.1in} {\bf Step 1}: Update ${\bf u}^{n+1}$ and
$s_\alpha^{n+1}$.
\begin{eqnarray}
\label{disc main step of 2nd order scheme}
\frac{s_\alpha^{n+1}-s_\alpha^{n}}{\Delta t}&=&T\left(\frac{s_\alpha^{n+1}+s_\alpha^{n}}{2}\right)+
\left(\bar{V}^*_\alpha-\theta_\alpha^{n+\frac{1}{2}}\bar{U}^*-T\left(s_\alpha^{n+\frac{1}{2}}\right)\right),\\
\rho\frac{\B{u}^{n+1}-\B{u}^n}{\Delta t}&=&-\nabla \bar{p}+\mu \nabla^2 \bar{\B{u}}+L_{n+\frac{1}{2}}
\left(\B{F}\left(\bar{s}_\alpha,\theta^{n+\frac{1}{2}}\right)\right),\\
\nabla^2 \bar{p}&=&\nabla \cdot L_{n+\frac{1}{2}}\left(\B{F}\left(\bar{s}_\alpha,\theta^{n+\frac{1}{2}}\right)\right) ,
\end{eqnarray}
The leading order terms, $T$ is given by
\begin{eqnarray}
\label{second order leading s}
\widehat{T}(\bar{s}_\al)&=&-\frac{S_b\Delta t}{4\rho\left(\D\min_\al s_\alpha^{n+\frac{1}{2}}\right)^2}\left(\frac{\left(\bar{\lambda}\D\min_\al s_\alpha^{n+\frac{1}{2}}\right)^2k^2+k^4}
{\sqrt{\left(\bar{\lambda}\D\min_\al s_\alpha^{n+\frac{1}{2}}\right)^2+k^2}}-|k|^3\right)\widehat{\bar{s}}_\al,\\
\end{eqnarray}
where $\D\bar{\lambda}^2=\frac{2\rho}{\mu\Delta t}$, and
$\B{u}^{*,n+1}$ is the intermediate velocity at $t^{n+1}$ which is
obtained by solving the unsteady Stokes equations implicitly but
with the forcing evaluated explicitly:
\begin{eqnarray}
\rho\frac{\B{u}^{*,n+1}-\B{u}^n}{\Delta t}&=&-\nabla \bar{p}^*+\mu \nabla^2 \bar{\B{u}}^*+L_{n+\frac{1}{2}}\left(\B{F}\left(s_\alpha^{n+\frac{1}{2}},\theta^{n+\frac{1}{2}}\right)\right),\\
\nabla^2 \bar{p}^*&=&\nabla \cdot L_{n+\frac{1}{2}}\left(\B{F}\left(s_\alpha^{n+\frac{1}{2}},\theta^{n+\frac{1}{2}}\right)\right) ,\\
\bar{V}^*&=&L^*_{n+\frac{1}{2}}\left(\bar{\B{u}}^*\right)\cdot \Bt^{n+\frac{1}{2}},\\
\bar{U}^*&=&L^*_{n+\frac{1}{2}}\left(\bar{\B{u}}^*\right)\cdot \B{n}^{n+\frac{1}{2}} .
\end{eqnarray}
and
\begin{eqnarray}
\bar{\B{u}}^*=\frac{\B{u}^{*,n+1}+\B{u}^{n}}{2}
\end{eqnarray}
\vspace{0.1in} {\bf Step 2}: After the ${\bf u}^{n+1}$ and
$s_\alpha^{n+1}$ are calculated in step 1, we update $\theta$ at
$t^{n+1}$ using the following semi-implicit scheme:
\begin{eqnarray}
\label{disc main step of 2nd order scheme theta}
\frac{\theta^{n+1}-\theta^{n}}{\Delta t}&=&\D
S\left(\frac{\theta^{n+1}+\theta^{n}}{2}\right)+\frac{1}{s_\alpha^{n+\frac{1}{2}}}\left(\bar{U}_\alpha+\bar{\theta}_\al
\bar{V}-S\left(\theta^{n+\frac{1}{2}}\right)\right),
\end{eqnarray}
where
\begin{eqnarray}
\bar{V}&=&L^*_{n+\frac{1}{2}}\left(\frac{\B{u}^{n+1}+\B{u}^{n}}{2}\right)\cdot \Bt^{n+\frac{1}{2}},\\
\bar{U}&=&L^*_{n+\frac{1}{2}}\left(\frac{\B{u}^{n+1}+\B{u}^{n}}{2}\right)\cdot
\B{n}^{n+\frac{1}{2}} .
\end{eqnarray}
and the leading order term is
\begin{eqnarray}
\label{second order leading t}
\widehat{S}(\bar{\theta})&=&-\frac{S_b\Delta
t\D\max_\al\left(s_\alpha^{n+\frac{1}{2}}-1\right)}{4\rho\left(\D\min_\al
s_\alpha^{n+\frac{1}{2}}\right)^3}\left(|k|^3-\frac{k^4}{\sqrt{\left(\bar{\lambda}\D\min_\al
s_\alpha^{n+\frac{1}{2}}\right)^2+k^2}}\right)\widehat{\bar{\theta}},
\end{eqnarray}
This completely defines our second order semi-implicit scheme.

\section{Numerical results}

In this section, we will perform a number of numerical experiments
to test the stability of our semi-implicit schemes for both the
steady and unsteady Stokes equations. We also compare the performance
of our semi-implicit schemes with the explicit scheme and the fully
implicit scheme. Our numerical results indicate convincingly that
our semi-implicit schemes has a much better stability
property that that of the explicit scheme. Moreover, the
computational cost of our semi-implicit schemes is comparable to
that of an explicit scheme. Thus our semi-implicit schemes offer
significant computational saving over the explicit scheme,
especially when the number of grid points is large.

\subsection{Model problem}

The test problem we use is one typically seen in the literature, in
which the immersed boundary is a closed loop initially in the shape
of an ellipse. We choose an ellipse initially aligned in the coordinate
directions with horizontal semi-axis $a=0.32$ and vertical semi-axis
$b=0.24$. The boundary can be parameterized as follows:
\begin{eqnarray}
\left\{
\begin{array}{c}
x(\al,0)=0.5+0.32\cos\al,\\
y(\al,0)=0.5+0.24\sin\al .
\end{array}
\right.
\end{eqnarray}
The fluid is initially at rest in a periodic domain
$\Omega=[0,1]\times[0,1]$. We use a periodic boundary condition
for the fluid flow. For the initial condition defined above, the
rest state of the boundary is a circle with radius $r=0.2$.
For the unsteady Stokes flow, the immersed boundary
with the above initial condition evolves as damped oscillations
around a circular equilibrium state. The area is conserved during
the time evolution since the flow is incompressible.
For the steady Stokes flow, the boundary converges
to the circular state without oscillations.

We use a uniform $N\times N$ grid to discretize the fluid domain,
$\Omega$. We choose $N_b=2N$ number of grid points to discretize
the immersed boundary so that there are approximately 2 immersed
boundary points per mesh width. We use the spectral method
to discretize the spatial derivatives both in the fluid
domain and along the immersed boundary. The leading order
singular integral is also discretized by the spectral method.

\subsection{Steady Stokes flow}

First, in order to reduce the number of parameters in our test problem, we write the equations in terms of the following
dimensionless variables to get the nondimensional model \cite{TP92},
\begin{eqnarray}
t'=\frac{t}{t_0},\quad \B{x}'=\frac{\B{x}}{L},\quad
\B{u}'=\frac{\B{u}t_0}{L},\quad p'=\frac{pt_0}{\mu}, \quad
\B{f}'=\frac{\B{f}Lt_0}{\mu} ,
\nonumber
\end{eqnarray}
where $L$ is the size of computational domain, $t_0$ is
characteristic time.

 Using these new variables, we have
\begin{eqnarray}
0&=&-\nabla p' + \triangle \mathbf{u}' + \mathbf{f}'(\mathbf{x'},t'), \\
0&=&\nabla \cdot \mathbf{u}' .
\end{eqnarray}
For the equations of the elastic boundary, the dimensionless variables are
\begin{eqnarray}
\B{X}'=\frac{\B{X}}{L}, \quad s_\al'=\frac{s_\al}{L},\quad \theta'=\theta,\quad \al'=\frac{\al}{L},
\quad T'=\frac{T}{S_b},\quad \B{F}'=\frac{\B{F}L}{S_b}, \quad
\Bt'=\Bt,\quad \B{n}'=\B{n} .\nonumber
\end{eqnarray}
Then the equations describe the interaction of the boundary and the fluid become
\begin{eqnarray}
U'&=&\B{u}'\left(\B{X'}(\alpha',t'),t'\right)\cdot \B{n'},\\
V'&=&\mathbf{u}'\left(\B{X'}(\alpha',t'),t'\right)\cdot \mbox{\boldmath$\tau'$},\\
s'_{\alpha, t'}&=&V'_{\alpha'}-\theta'_{\alpha'} \,U',\\
\theta'_{t'}&=&\frac{1}{s'_\al}\left(U'_{\alpha'}+
V'\,\theta'_{\alpha'}\right),
\end{eqnarray}
where
\begin{eqnarray}
\displaystyle\mathbf{f'}(\mathbf{x'},t')&=&\frac{S_bt_0}{\mu L}\int_0^{L_b/L}\mathbf{F'}(\alpha',t')\delta
(\mathbf{x'-X'}(\alpha',t'))d\alpha' ,\\
\B{u}'(\B{X}'(\al',t'),t')&=&\int_\Omega \B{u}'(\B{x}',t')\delta(\B{x}'-\B{X}'(\al',t'))d\B{x}' .
\end{eqnarray}
There are two nondimensional parameters in this problem:
\begin{eqnarray}
\frac{S_bt_0}{\mu L}, \quad \frac{L_b}{L}.\nonumber
\end{eqnarray}
If we let $\D t_0=\frac{\mu L}{S_b}$, then the only parameter in this dimensionless model is $L_b/L$
which is fixed in our test problem. So we can always fix $S_b=\mu=1$ in our numerical study.

\begin{figure}\center
\includegraphics[width=0.4\textwidth]{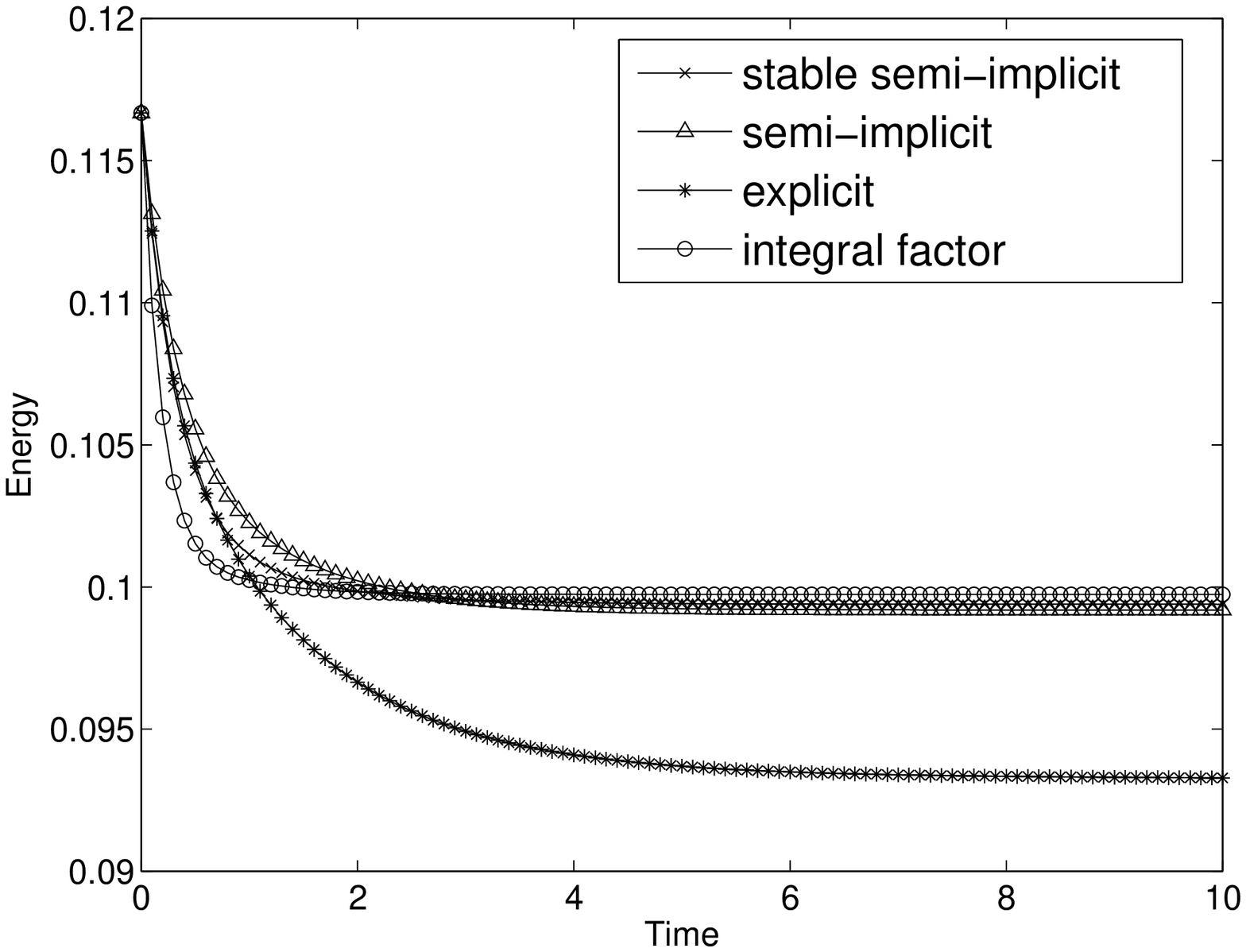}
\quad
\includegraphics[width=0.4\textwidth]{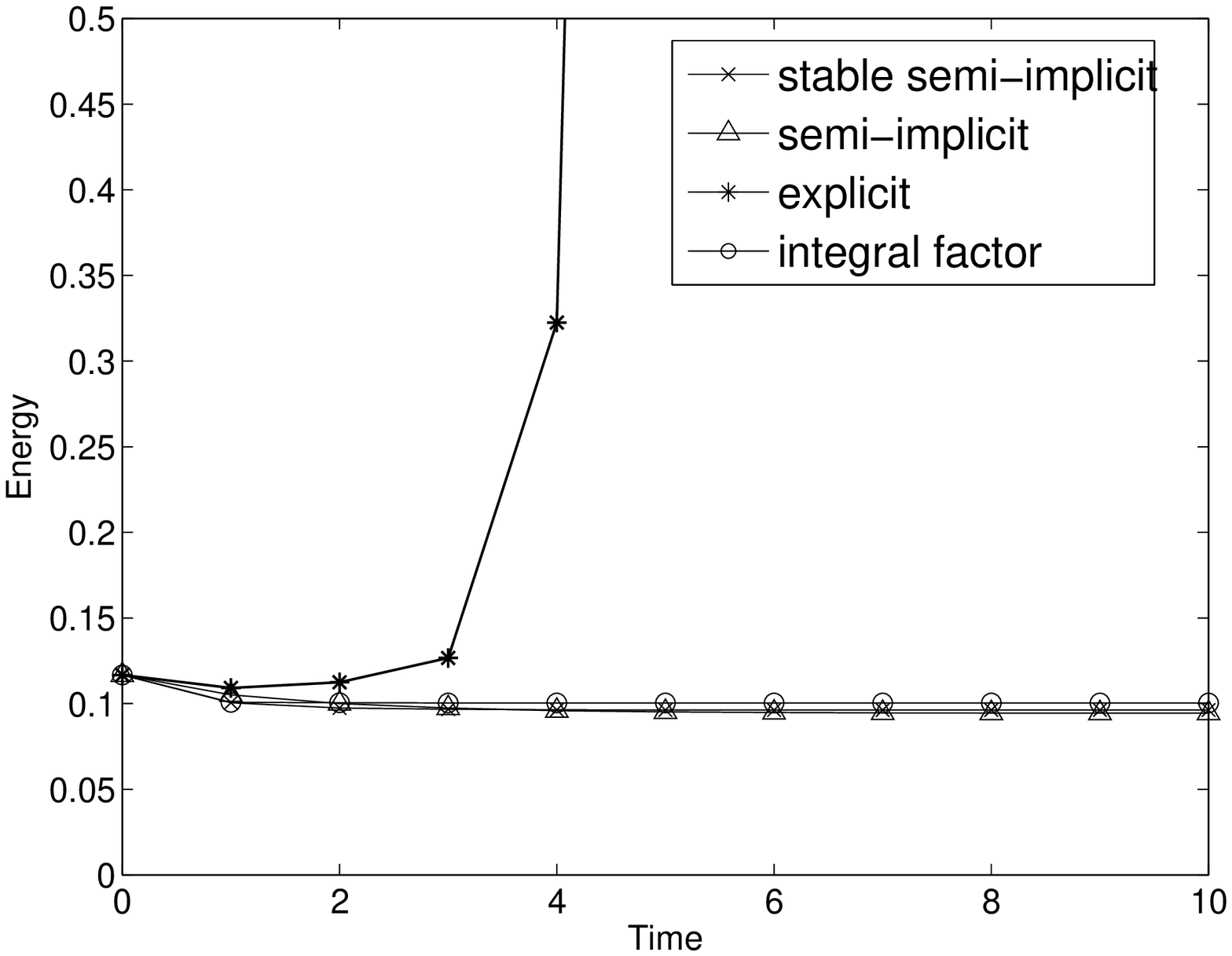}
\caption{Energy of the system for four different schemes. $N=128$,
$S_b =1$, $\mu =1$. Left one: $\Delta t=0.1$; Right: $\Delta t=1$.
Here the legend ``stable semi-implicit'' stands for the
unconditionally stable semi-implicit scheme, ``semi-implicit'' for
the semi-implicit scheme of first kind, and ``integral factor'' for
the semi-implicit scheme
 based on the integral factor method.}
\label{steady 0.1 and 1}
\end{figure}

\begin{figure}
\center
\includegraphics[width=0.5\textwidth]{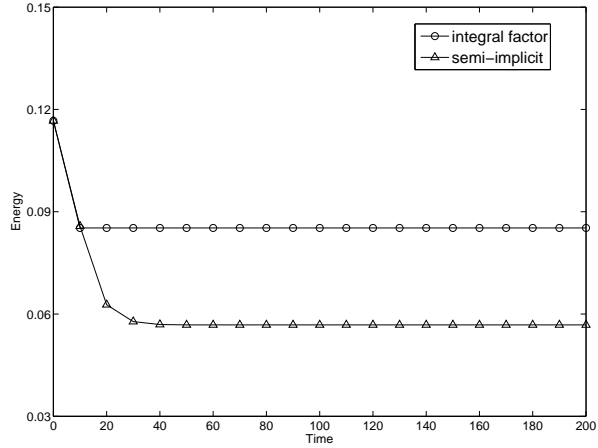}
\caption{Energy for the semi-implicit scheme of first kind (labeled
as ``semi-implicit'') and the semi-implicit scheme based on the
integral factor method (labeled as ``integral factor''). $\Delta
t=10$, $N=128$, $S_b =1$, $\mu =1$.} \label{energy steady 10}
\end{figure}

\begin{figure}
\center
\includegraphics[width=0.4\textwidth]{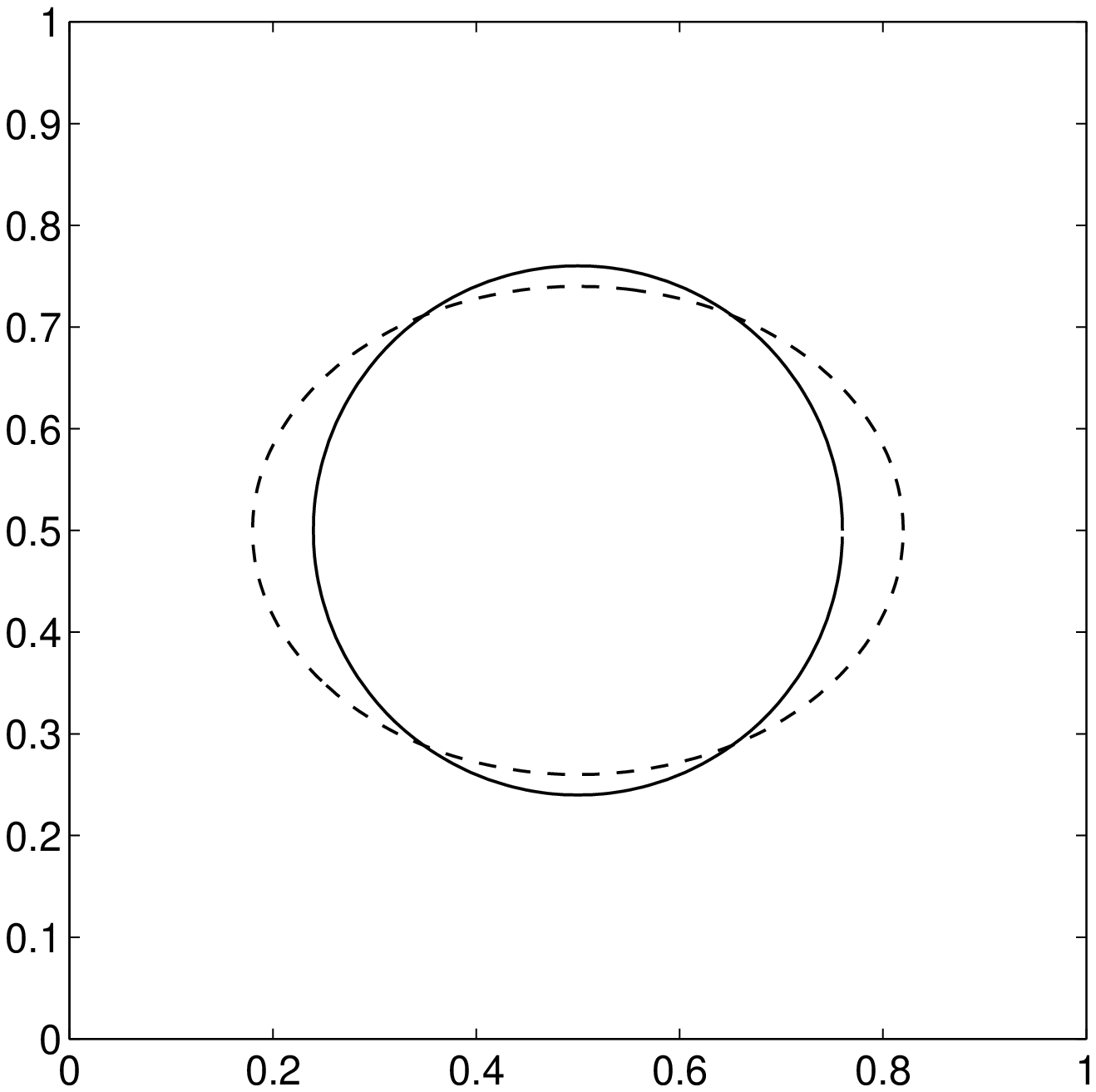}
\quad
\includegraphics[width=0.4\textwidth]{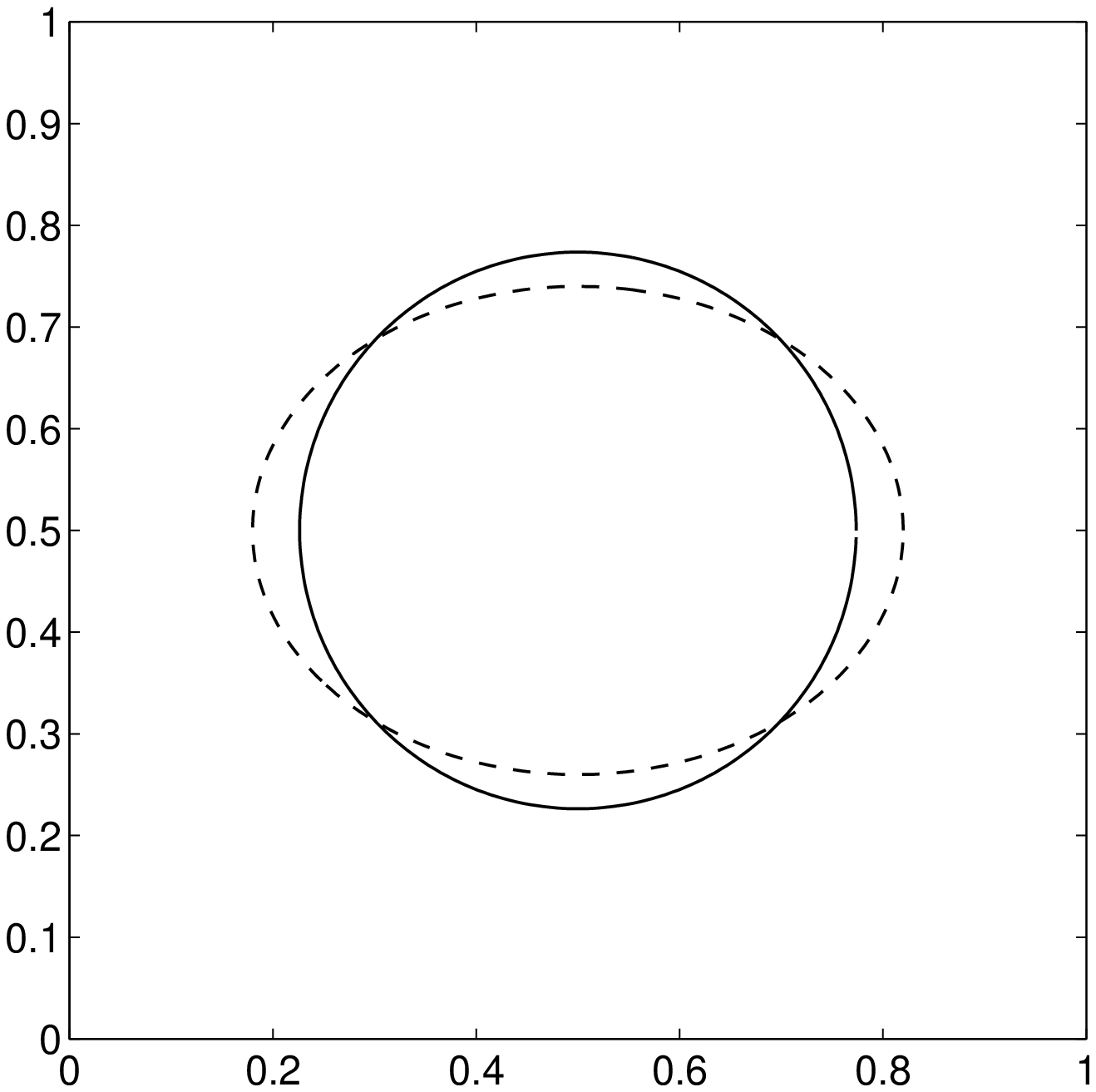}
\caption{Dashed line: the initial boundary configuration; Solid
line: the boundary configuration after 20 time steps with $\Delta
t=10$, $N=128$, $S_b =1$, $\mu =1$. Left: semi-implicit ; Right:
integral factor method.} \label{conf steady 10}
\end{figure}

The stability analysis in the steady Stokes flow suggests us to use
the total energy as a criterion to test the stability of different
numerical methods. For the steady Stokes equations, the total
energy is equal to the potential energy. In Fig \ref{steady 0.1 and 1},
we show that the energy for four different numerical methods: the
explicit scheme, the semi-implicit scheme of first kind,
the 4th order semi-implicit scheme using the integral
factor method and the unconditionally stable semi-implicit scheme.
In this figure and the subsequent figures, we use the
the legend ``semi-implicit'' to denote the semi-implicit scheme
 of first kind, and the legend
``integral factor'' to denote the semi-implicit scheme based on the integral factor method. We use two different time
steps, 0.1 and 1, respectively.
When $\Delta t=0.1$, all the four methods are stable. They
give almost identical results. When $\Delta t=1$, the explicit scheme becomes unstable
, but all the semi-implicit schemes are
stable. In fact, all the semi-implicit schemes remain stable with
much larger time steps. In Fig \ref{energy steady 10}, we plot
the energy of the system for semi-implicit schemes of first kind and
the semi-implicit scheme based on the integral factor method
with $\Delta t=10$. Fig \ref{conf steady 10} shows the
configuration obtained by the two semi-implicit schemes
at the final time with $\Delta t =10$. They both remain as a
circle, but lose some area compared with the original state.


Next, we compare the performance of our semi-implicit schemes with the
explicit and fully implicit schemes. The fully implicit scheme we use
here was originally proposed by Tu and Peskin in \cite{TP92}.
In order to make a fair comparison, we run the implicit schemes
(semi-implicit and fully implicit) with a time step small enough to
make sure that the computational results have a reasonable accuracy. We take
$\Delta t=4$ for the fully implicit and the semi-implicit schemes.
With this time step, the area loss is less than
5\%.
For the explicit scheme, we take $\Delta
t=1/4,1/8,1/16,1/32$ which corresponds to $N=64,128,256,512$
respectively. These time steps are
the largest possible to keep the stability of the explicit scheme.
The time we compute is $T=20$. The result is shown in Table
\ref{steady com time}.
\begin{table}
\begin{center}
\begin{tabular}{|c||c|c|c|c|c|}\hline
 $N$& exp& s,i 1& s,i 2 & s,s,i&f,i \\ \hline
 64 & 1 & 0.4 & 2 & 7 & 9 \\ \cline{1-6}
128 & 5 & 0.7 & 3 & 30 & 39 \\ \cline{1-6}
256 & 30 & 1.3 & 7 & 139 & 206 \\ \cline{1-6}
512 & 344 & 4.2 & 19 & 611 & 1200 \\
\hline
\end{tabular}
\end{center}
\caption{Execution time for each computation in seconds. The legends
are defined as follows: ``exp'' stands for the explicit scheme,
``s,i1'' the semi-implicit method of first kind, ``s,i2'' the
semi-implicit scheme of the second kind , ``s,s,i'' the
unconditionally stable semi-implicit method, and ``f,i'' the fully
implicit scheme . $N$ is the number of grid points along each
dimension.} \label{steady com time}
\end{table}
From this comparison, we can see that the performance of our
semi-implicit schemes is much better than the explicit, the
fully implicit scheme, and the unconditionally stable semi-implicit
scheme in all cases. As we can see, the larger the
number of the spatial grid points is, the more computational saving
we would get using our semi-implicit schemes. Even for a modest grid
size, our semi-implicit schemes still give a significant
computational saving compared with the explicit or the fully
implicit scheme. It is interesting to note that
although the computational cost of the unconditionally stable
semi-implicit scheme (labeled as s,s,i) is faster than
the fully implicit method, the computational cost of the
unconditionally stable semi-implicit method is still more
expensive than the explicit scheme. This makes the unconditionally
stable semi-implicit scheme not very practical.

\subsection{Unsteady Stokes flow}
We can also get the nondimensional model for unsteady stokes flow. Similar as the steady
stokes case, we define the following dimensionless variables:
\begin{eqnarray}
t'=\frac{t}{t_0},\quad \B{x}'=\frac{\B{x}}{L},\quad \B{u}'=\frac{\B{u}t_0}{L},\quad p'=\frac{pt_0}{\mu},\quad \B{f}'=\frac{\B{f}Lt_0}{\mu} .
\nonumber
\end{eqnarray}
where $L$ is the size of computational domain, $t_0$ is
characteristic time.
Using these new variables, we have
\begin{eqnarray}
\frac{\p \B{u}'}{\p t'}&=&\frac{\mu t_0}{\rho L^2}\left(-\nabla p' + \triangle \mathbf{u}' + \mathbf{f}'(\mathbf{x'},t')\right), \\
0&=&\nabla \cdot \mathbf{u}'.
\end{eqnarray}
For the equations of the elastic boundary, the dimensionless variables are
\begin{eqnarray}
\B{X}'=\frac{\B{X}}{L},\quad s_\al'=\frac{s_\al}{L},\quad \theta'=\theta,\quad \al'=\frac{\al}{L},\quad
T'=\frac{T}{S_b},\quad \B{F}'=\frac{\B{F}L}{S_b}, \quad
\Bt'=\Bt,\quad \B{n}'=\B{n} .\nonumber
\end{eqnarray}
Then the equations describe the interaction of the boundary and the fluid become
\begin{eqnarray}
U'&=&\B{u'(X'}(\alpha',t'),t')\cdot \B{n'},\\
V'&=&\mathbf{u'(X'}(\alpha',t'),t')\cdot \mbox{\boldmath$\tau'$},\\
s'_{\alpha, t'}&=&V'_{\alpha'}-\theta'_{\alpha'} U',\\
\theta'_{t'}&=&\frac{1}{s'_\al}\left(U'_{\alpha'}+
V'\theta'_{\alpha'}\right),
\end{eqnarray}
where
\begin{eqnarray}
\displaystyle\mathbf{f'}(\mathbf{x'},t')&=&\frac{S_bt_0}{\mu L}\int_0^{L_b/L}\mathbf{F'}(\alpha',t')\delta
(\mathbf{x'-X'}(\alpha',t'))d\alpha' ,\\
\B{u}'(\B{X}'(\al',t'),t')&=&\int_\Omega \B{u}'(\B{x}',t')\delta(\B{x}'-\B{X}'(\al',t'))d\B{x}' .
\end{eqnarray}
From the nondimensional analysis, we can see that there are three nondimensional parameters in this problem:
\begin{eqnarray}
\frac{S_bt_0}{\mu L}, \quad \frac{\mu t_0}{\rho L^2}, \quad \frac{L_b}{L}.\nonumber
\end{eqnarray}
If we let $\D t_0=\frac{\mu L}{S_b}$, then the parameters left in this dimensionless model is $\D\frac{\mu t_0}{\rho L^2}=\frac{\mu^2}{\rho L S_b}$ and $\D\frac{L_b}{L}$. $L_b$ and $L$ are fixed in our test problem only depends on the initial condition. So
$\D\frac{\mu^2}{\rho L S_b}$ is the only parameter in our test model. For this reason, we always fix the elastic coefficient
$S_b$ to 1, but vary $\mu$.

In our computations, we use the following parameter values:
\begin{eqnarray}
\rho=1, S_b=1, \mu=0.1,0.01,0.005. \nonumber
\end{eqnarray}
We vary the number of the spatial grid points along each
dimension in the following fashion:
\begin{eqnarray}
N=64,128,256,512.\nonumber
\end{eqnarray}

The criterion that we use to check whether one scheme is stable or not
is that the total energy of the system is non-increasing and the
boundary configuration lies within the computational domain.

\begin{figure}
\center
\includegraphics[width=0.4\textwidth]{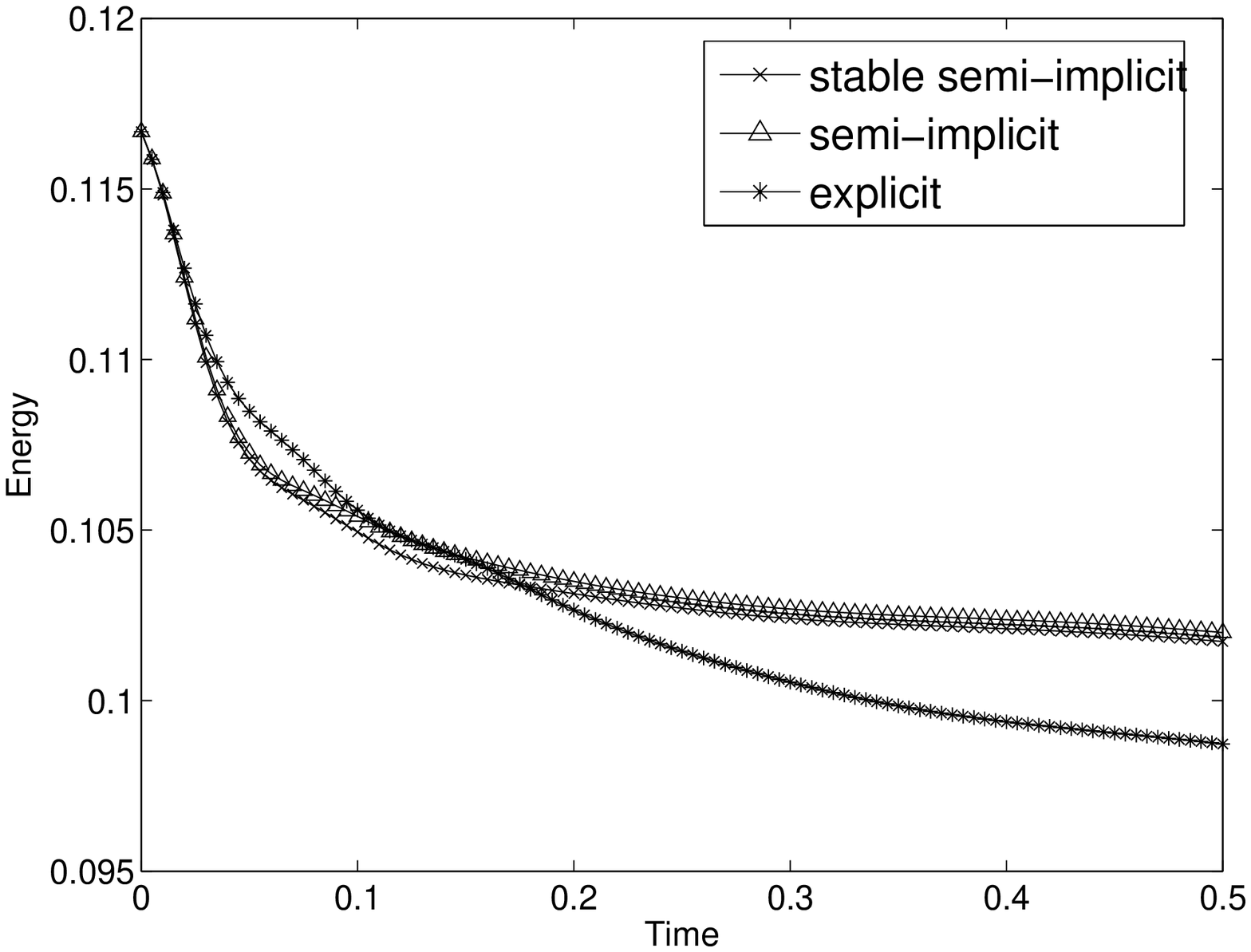}
\quad
\includegraphics[width=0.4\textwidth]{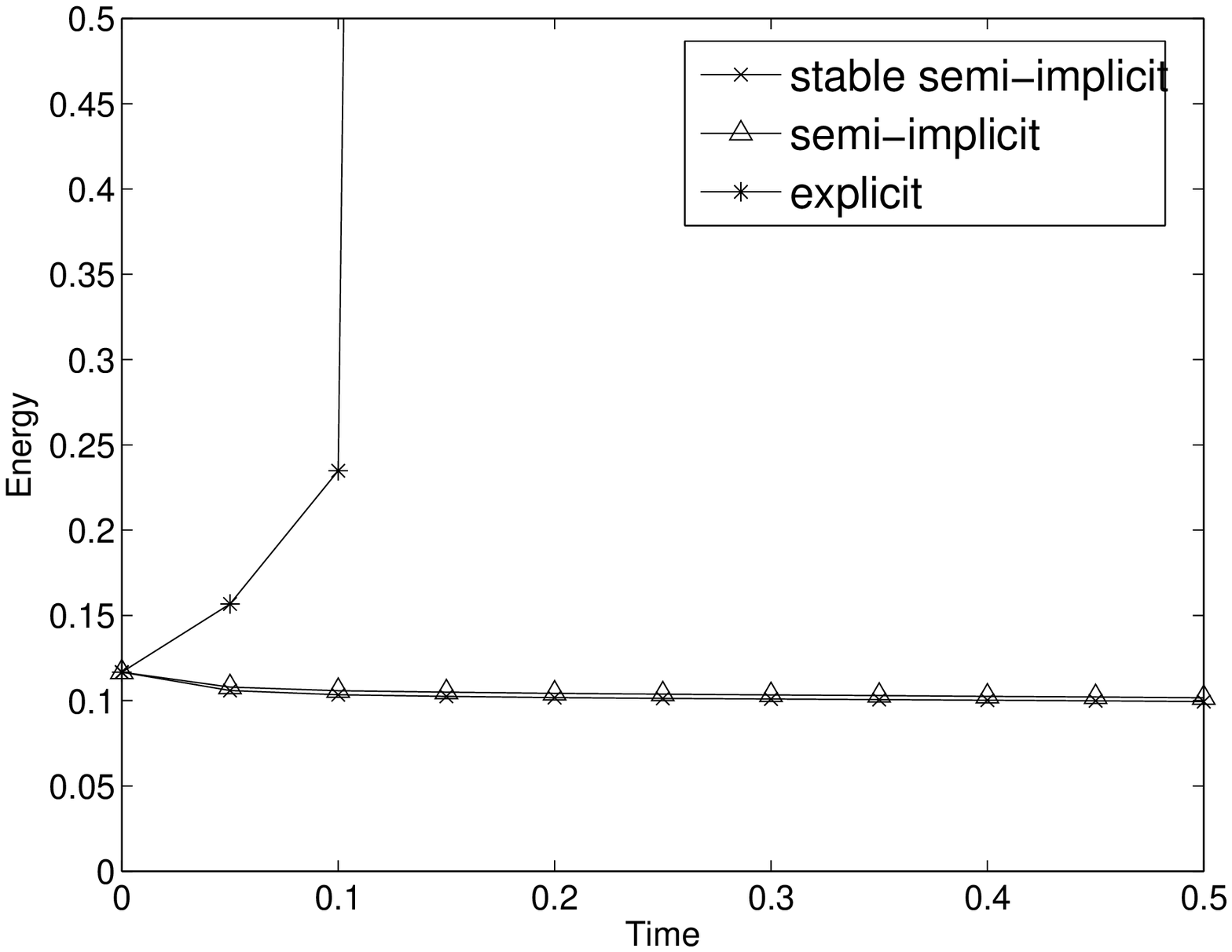}
\caption{Total energy of the unsteady Stokes system for
different schemes with two different timesteps.
$N=128$, $S_b =1$, $\mu =0.01$.
Left: $\Delta t=0.005$; Right: $\Delta t=0.05$. The
legend ``semi-implicit'' stands for the solution
obtained by the semi-implicit scheme of first kind,
``stable semi-implicit'' the unconditionally stable
semi-implicit method.}
\label{unsteady 0.005 and 0.05}
\end{figure}

\begin{figure}
\center
\includegraphics[width=0.4\textwidth]{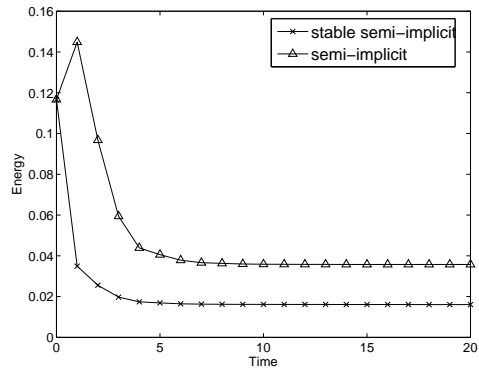}

\caption{Total energy of the system for two semi-implicit schemes. Here
``stable semi-implicit'' stands for the
unconditionally stable semi-implicit scheme,
and ``semi-implicit'' the semi-implicit scheme of first kind,
 $\Delta t=1$, $N=128$, $S_b =1$, $\mu =0.01$.}
\label{unsteady 1}
\end{figure}

Next, we perform some numerical experiments to test the stability of
our semi-implicit schemes for the unsteady Stokes flow.
Fig. \ref{unsteady 0.005 and 0.05} shows that the energy obtained
by the explicit scheme and the semi-implicit scheme of the first
kind. We take two different timesteps, 0.005 and 0.05.
With $\Delta t=0.005$, the explicit and semi-implicit schemes are
all stable, and they give nearly identical results. With
$\Delta t=0.05$, the
explicit scheme becomes unstable, but the semi-implicit schemes
remain stable. Even if we increase the timestep to
$\Delta t = 1$, the semi-implicit methods are still stable, as
we can see from Fig. \ref{unsteady 1}. For the semi-implicit
scheme of the first kind, we have used the Small Scale Decomposition
and further simplification of the singular integral kernel. Therefore,
the total energy in our semi-implicit scheme is not guaranteed
to decrease monotonically in time. Nonetheless,
we observe that the total energy still decreases in time as
is the case for the unconditionally stable semi-implicit scheme.
In Fig. \ref{unsteady 1}, we also plot the boundary configuration
at the final time step, which is an approximate circle.

\begin{figure}
\center
\includegraphics[width=0.4\textwidth]{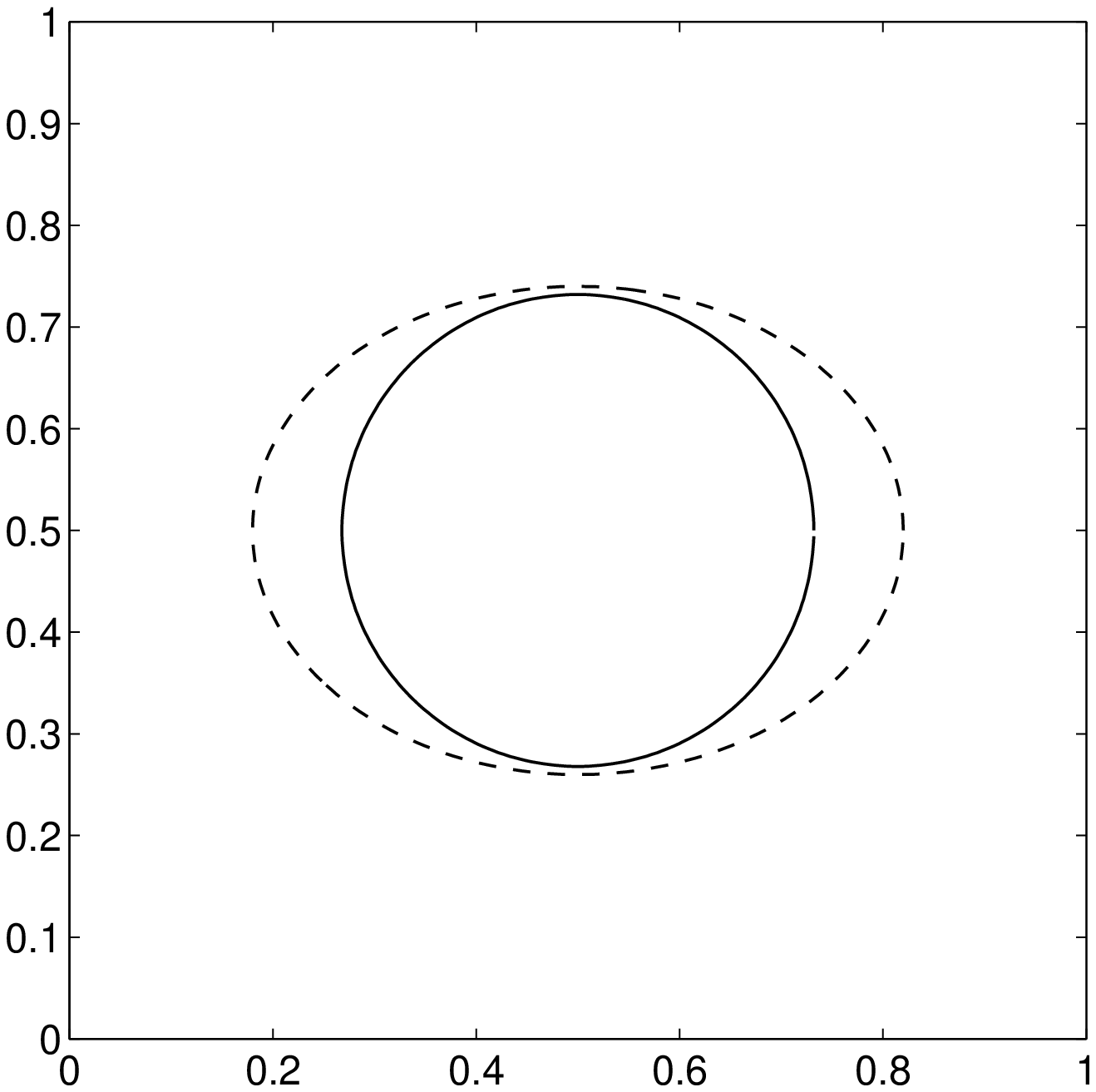}
\quad
\includegraphics[width=0.4\textwidth]{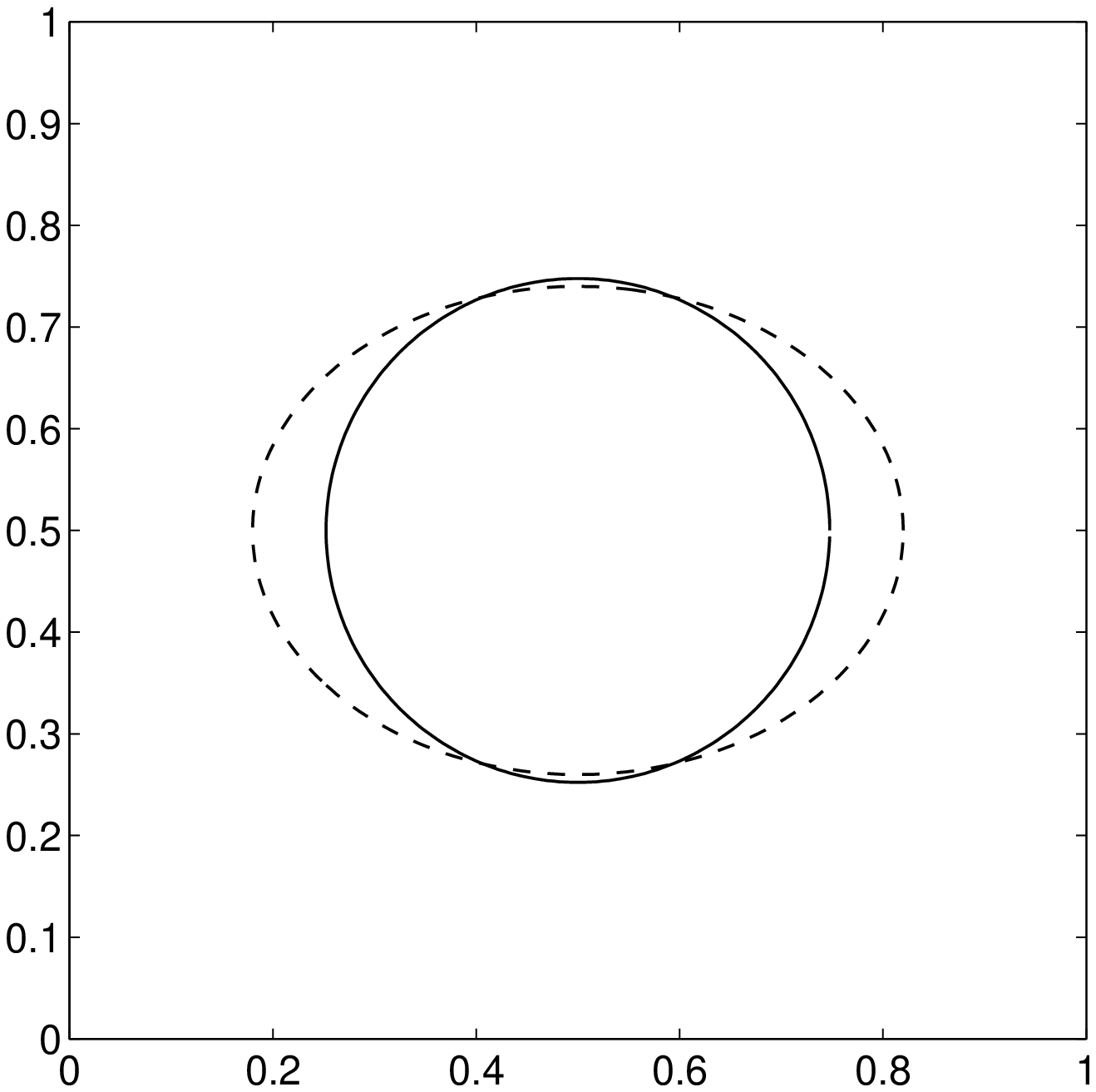}
\caption{Dashed line: the initial boundary configuration; Solid line: the
boundary configuration after 20 time steps with $\Delta t=1$,
$N=128$, $S_b =1$, $\mu =0.01$. Left: the unconditionally stable
semi-implicit scheme; Right: the semi-implicit scheme of the first kind.}
\label{unsteady 1 con}
\end{figure}

We remark that the semi-implicit scheme is not unconditionally
stable, although its stability is much better than the explicit scheme.
This is due to the fact that we have used the Small Scale Decomposition
and further approximation of the leading order singular integral
operator to simplify the computation of the implicit solution. As we
mentioned before, the Small Scale Decomposition captures only the
high frequency contribution to the stiffness, but it does not remove
the stiffness of the system induced by the low frequency components
of the solution. Thus there is still some mild timestep stability
constraint for the semi-implicit scheme. The
time step has a mild dependence on the elastic coefficient $S_b$
and the viscous coefficient $\mu$. On the other hand, our numerical
study shows that the time step for the semi-implicit scheme is
independent on the meshsize.

We also compare the performance of our semi-implicit schemes with
the explicit scheme. We do not compare the performance of our
semi-implicit schemes with the fully implicit scheme here because
the fully implicit scheme is quite expensive and is not competitive
with the explicit scheme. In order to keep the area loss is no more
than 5\%, we take $\Delta t=\frac{1}{4}$ for all of the
semi-implicit schemes . For the explicit scheme, we take $\Delta
t=1/64,1/128,1/256,1/512$ which correspond to the spatial mesh sizes
$N=64,128,256,512$ respectively, when $\mu=0.05$. When $\mu=0.01 $
and $0.005$, the time step is set to be $\Delta
t=1/128,1/256,1/512,1/1024$ and $t=1/256,1/512,1/1024, 1/2048$.
These time steps are the largest ones we can take to keep the
stability. We compute the solution up to $T=2$. The results are
documented in Table \ref{unsteady com time}. We can clearly see that
the semi-implicit scheme of the first kind gives a significant
improvement over the explicit scheme. The cost for the semi-implicit
scheme of the second kind is higher than that for the semi-implicit
scheme of the first kind. This is because we need to solve for a
linear system at each time step for the semi-implicit scheme of the
second kind. The cost increases as the number of the spatial grid
points increases. The semi-implicit scheme of the second kind and
the unconditionally stable semi-implicit scheme both need to solve a
linear system at each time step. Their complexity are same, both are
$O(N_b^2)$. But for the unconditionally stable semi-implicit scheme,
the scaling constant in front of $N_b^2$ is much larger than the
semi-implicit scheme of the second kind.
The reason is that the cost of computing the
coefficient matrix of the linear system for the
unconditionally stable semi-implicit scheme is much higher.
As we can see from Table 2, the unconditionally
stable semi-implicit scheme (labeled as s,s,i) is still quite
expensive compared with our semi-implicit schemes that use
the Small Scale Decomposition. For $\mu=0.05$, the
unconditionally stable semi-implicit scheme is even more
expensive compared with the explicit scheme. Although the
unconditionally stable semi-implicit scheme is slightly faster
than the explicit scheme for smaller $\mu$, the semi-implicit scheme
(labeled as s,i,1) which uses SSD to further simply the
singular integral kernel gives a much more efficient algorithm.
It gives a factor of 242 times speed-up over the explicit
scheme in the case of $\mu=0.005$ and $N=512$.

\begin{table}
\begin{center}
\begin{tabular}{|c||c|c|c|c|c|c|c|c|c|c|c|c|}\hline
 $N$&\multicolumn{4}{c|}{$\mu=0.05$}&\multicolumn{4}{c|}{$\mu=0.01$}&
\multicolumn{4}{c|}{$\mu=0.005$}\\ \cline{2-13} & exp& s,i 1& s,i
2&s,s,i& exp& s,i 1& s,i 2&s,s,i& exp& s,i 1& s,i 2&s,s,i  \\
\hline
 64 & 1.8 & 0.5 & 4 & 11 & 3.3 & 0.5 & 4 & 12 &6.6 & 0.5 & 4 &12  \\ \hline
128 & 9 & 1 & 10&  48 & 18& 0.9 & 10 & 47 &  35 &0.9 & 10 &48 \\ \hline
256 & 58 & 2.4 & 25 & 229 & 116 & 2.4 & 25 & 228 & 236 & 2.4 & 25 & 226 \\
\hline
512 & 738 & 12 & 99 & 980 & 1461 & 12 & 98 & 982& 2910 & 12 & 98 & 977 \\
\hline
\end{tabular}
\end{center}
\caption{Execution times for each computation in seconds. The legends
are defined as follows: ``exp'' stands for the
explicit scheme, ``s,i1'' the semi-implicit scheme of the first kind,
``s,i2'' the semi-implicit scheme of the second kind,
``s,s,i'' the unconditionally stable semi-implicit scheme. }
\label{unsteady com time}
\end{table}

\subsection{Second order semi-implicit scheme for the unsteady Stokes flow}

\begin{figure}
\center
\includegraphics[width=0.4\textwidth]{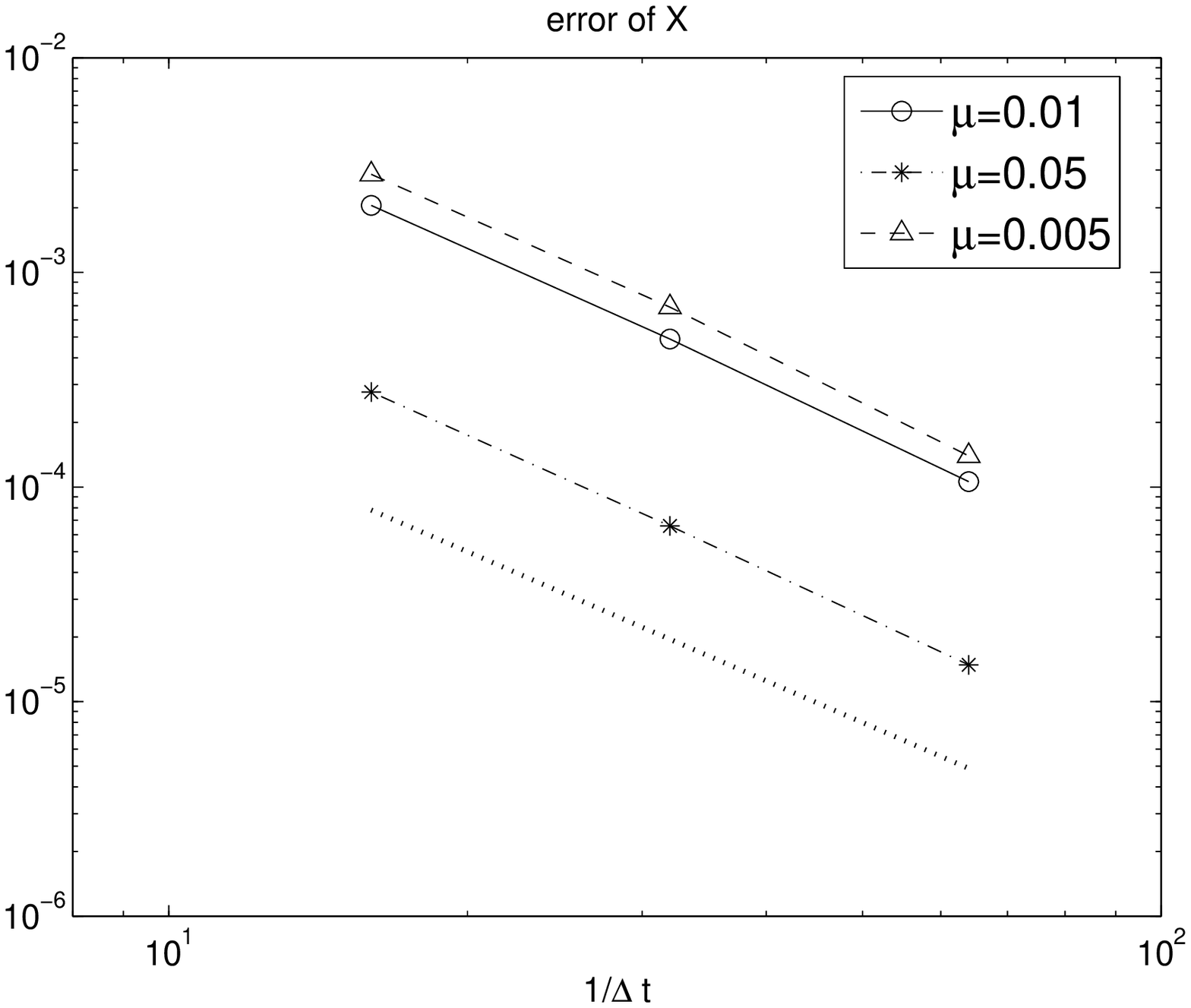}
\quad
\includegraphics[width=0.4\textwidth]{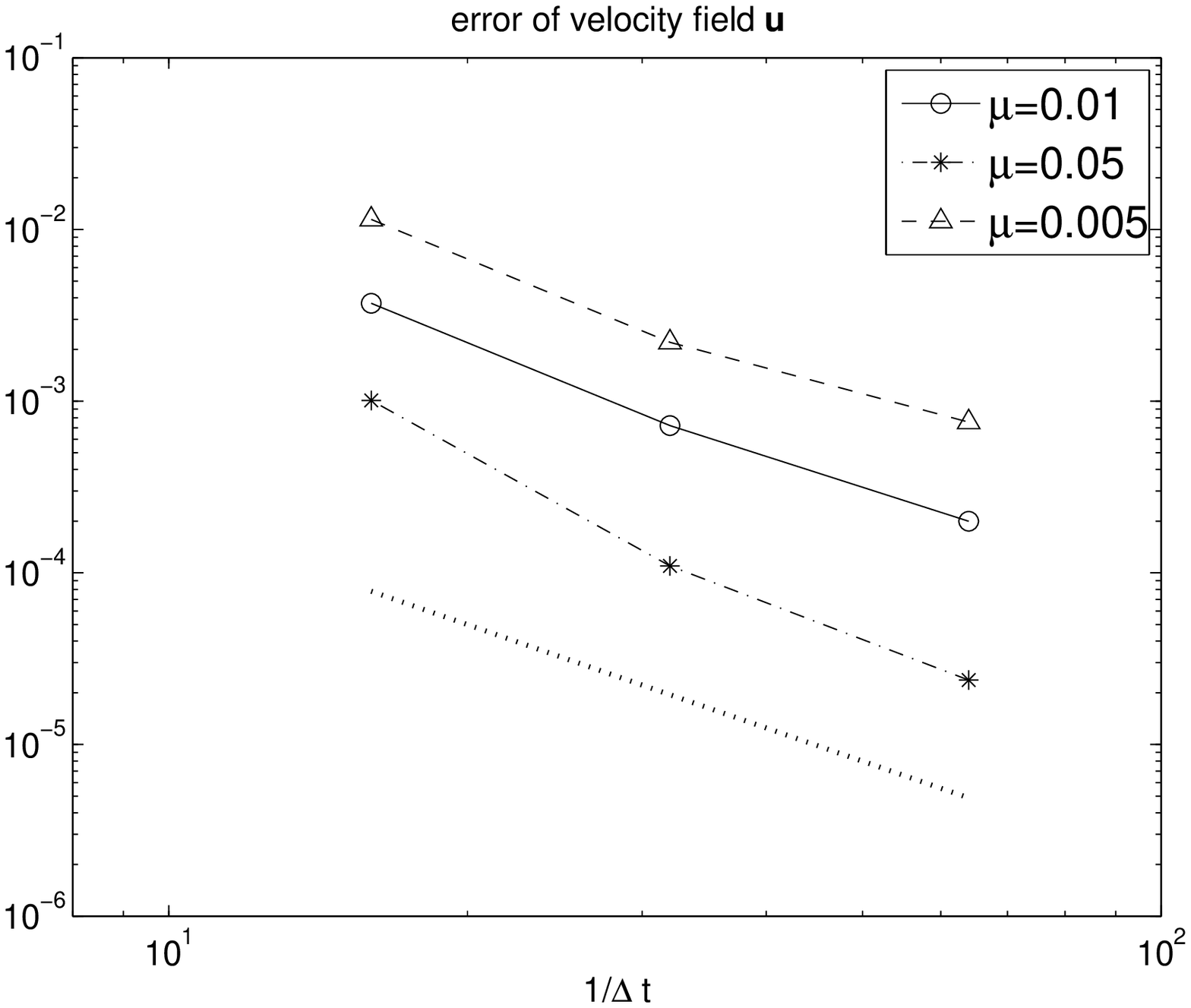}
\caption{Plot of $l^2$ errors in time at time $T=1$ for the second order
semi-implicit scheme of the first kind. We choose $S_b = 1$ and
$N=256$ in all computations
except in the case of $\mu=0.005$ where $N$ is increased to $512$.
The line at the bottom of each graph is a reference line
which corresponds to the second order convergence rate. }
\label{unsteady converge}
\end{figure}

In this subsection, we perform numerical experiments to test the
convergence rate and the stability property of our second order
semi-implicit scheme. To check the convergence rate in time, we set
$N=256$ and vary the time step in powers of 2 from
$\frac{1}{16}$ to $\frac{1}{128}$. When $\mu=0.005$, the solution
becomes more singular. In order to fully resolve the spatial
solution, we increase the spatial resolution to $N=512$.
Following \cite{MP07}, we compute the time discretization
error at time $t$ as follows:
\begin{eqnarray}
e_T(u;\Delta t)=\|u(T;\Delta t)-u(T;\Delta t/2)\|_{l^2}.
\end{eqnarray}
For a vector field $\B{u}(\B{x})=(u_1(\B{x}),u_2(\B{x}))$ defined on the
Cartesian grid with $x_i = i h$, $y_j = jh$, the discrete $l^2$ norm is
defined as follows
\begin{eqnarray}
\|\B{u}\|_{l^2}=\left(\sum_{i,j}\left(u_1^2(x_i,y_j)+u_2^2(x_i,y_j)\right)h^2\right)^{\frac{1}{2}}.
\end{eqnarray}
Similarly, the discrete $l^2$ norm for a vector field
$\B{w}(\al)=(w_1(\al),w_2(\al))$ defined on the interface $\Gamma$ is defined
below:
\begin{eqnarray}
\|\B{w}\|_{l^2}=
\left(\sum_{i}\left(w_1^2(\al_i)+w_2^2(\al_i)\right)
\Delta \al\right)^{\frac{1}{2}}.
\end{eqnarray}
We compute the solution up to $T=1$ and evaluate the convergence rate
based on the numerical solution at $T=1$ with different temporal resolutions.
The results are shown in Fig. \ref{unsteady converge} and Table
\ref{convergence rate}. As we can see, the convergence rate is
approximately second order.

\begin{table}
\begin{center}
\begin{tabular}{|c||c|c|}\hline
 $\mu$& convergence rate of $\B{X}$ & convergence rate of $\B{u}$  \\ \hline
 0.05 & 2.11 & 2.70  \\ \hline
0.01 & 2.13 & 2.10  \\ \hline $0.005^*$ & 2.17 & 1.96
\\ \hline
\end{tabular}
\end{center}
\caption{Convergence rates for $\B{X}$ and $\B{u}$ fitted from the data shown
in Fig \ref{unsteady converge}. the case * is computed using a refined mesh
$512\times 512$.}
\label{convergence rate}
\end{table}

\begin{figure}
\center
\includegraphics[width=0.4\textwidth]{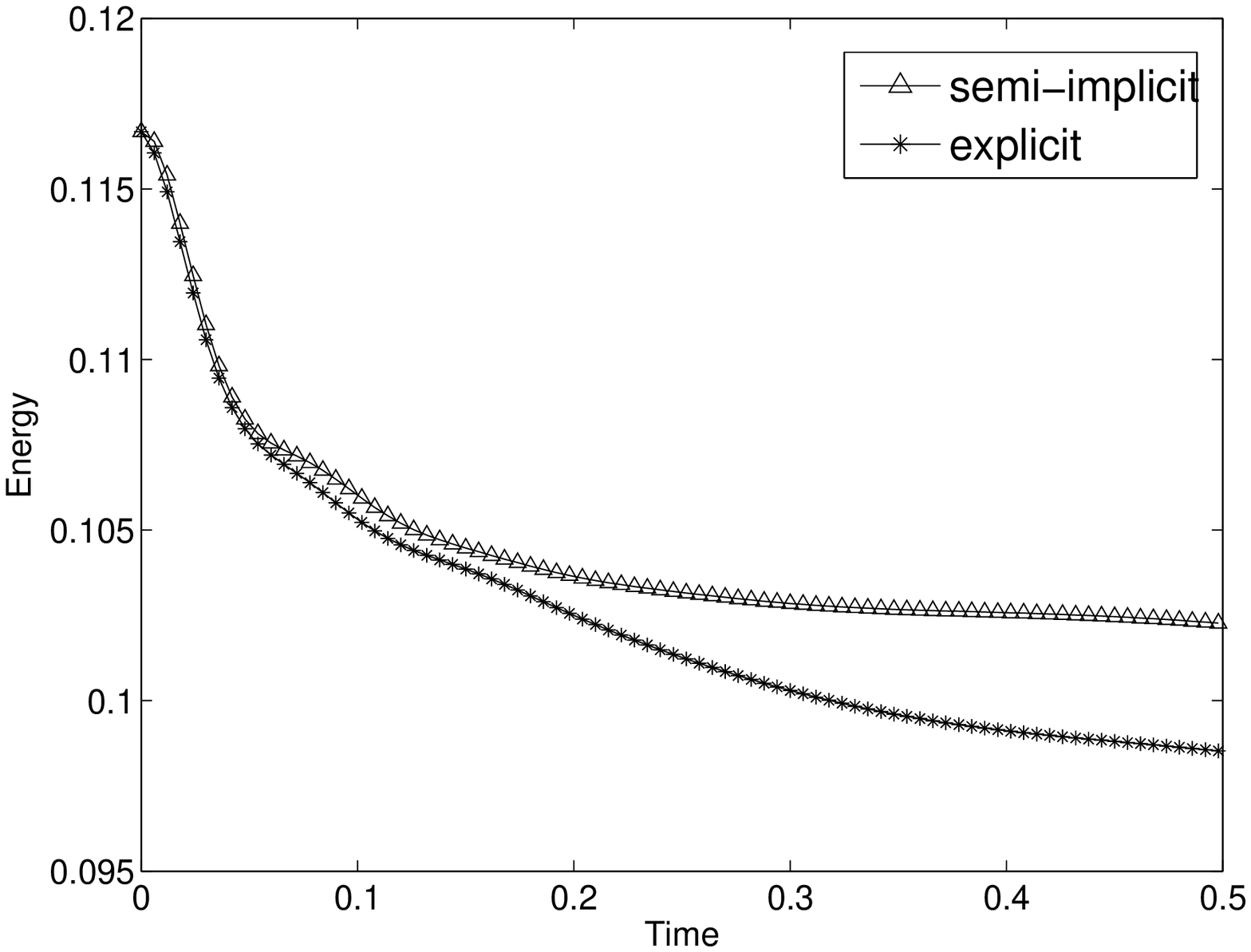}
\quad
\includegraphics[width=0.4\textwidth]{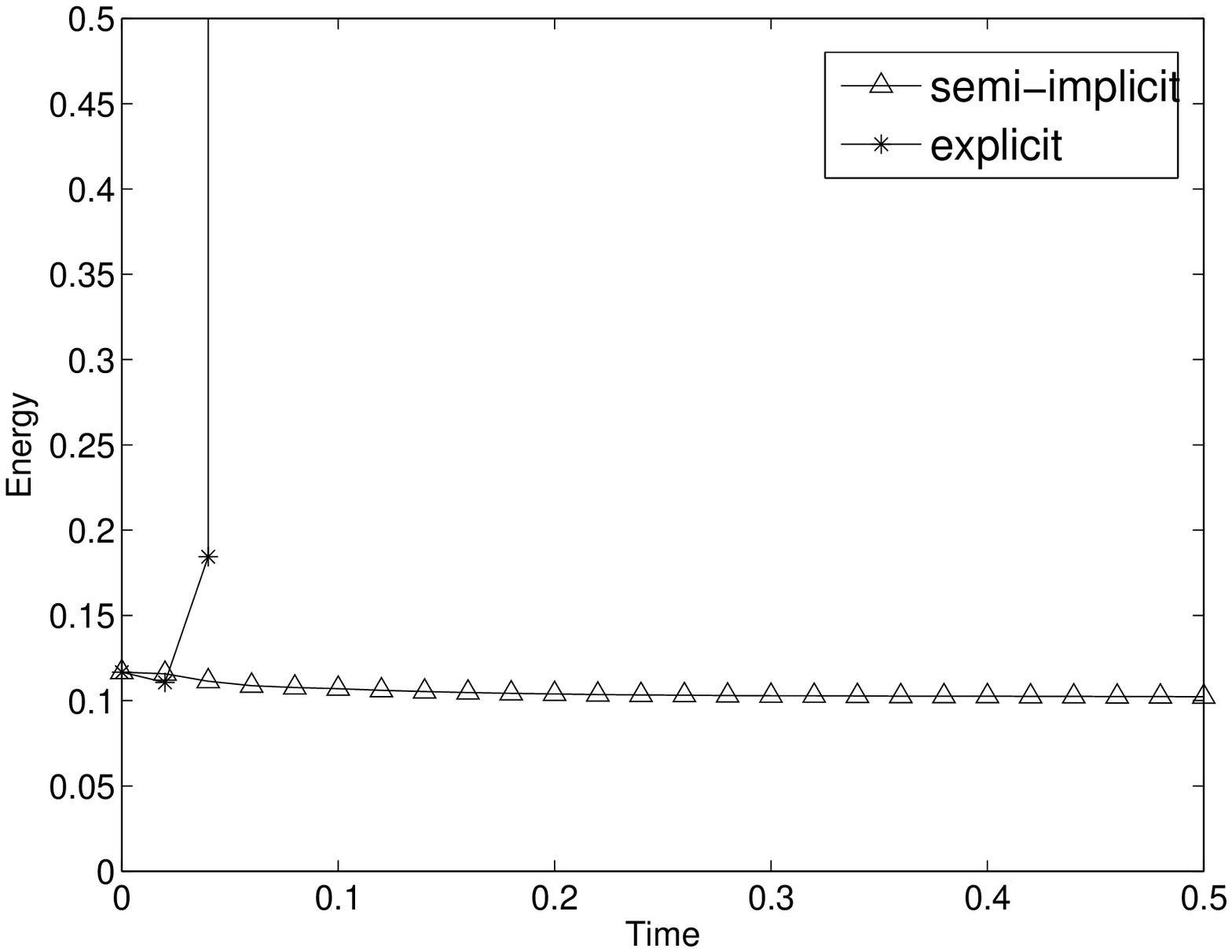}
\caption{Total energy of the unsteady Stokes system for the
second order semi-implicit scheme and the second order explicit scheme.
$N=128$, $S_b =1$, $\mu =0.01$.
Left: $\Delta t=0.002$; Right: $\Delta t=0.02$.}
\label{2nd unsteady 0.002 and 0.02}
\end{figure}

Next we check the stability of the second order semi-implicit scheme.
Fig \ref{2nd unsteady 0.002 and 0.02} shows that the total energy for the
second order
explicit and second order semi-implicit schemes with different timesteps
0.002 and 0.02. We choose the same second order explicit scheme that was used
in \cite{LP00}. With $\Delta t=0.002$, both the explicit and semi-implicit
schemes are stable and they give nearly identical results. With
$\Delta t=0.02$, the explicit scheme becomes unstable, but the semi-implicit
scheme is still stable. The stability restriction of the semi-implicit scheme
is far less severe than the corresponding explicit scheme.
Our numerical study also shows that the time step for the semi-implicit
scheme is independent on the meshsize while the explicit scheme requires
finer time step as the spatial mesh is refined.

Finally, we compare the performance of our second order
semi-implicit scheme with that of the second order explicit scheme.
As before, in order to keep the accuracy
with 5\%, we take $\Delta t=\frac{1}{4}$ for our
semi-implicit schemes. For the explicit scheme,
we take $\Delta t=1/128,1/256,1/512,1/1024$ which correspond to the
spatial mesh sizes $N=64,128,256,512$ respectively, when
$\mu=0.05$. When $\mu=0.01$ and $0.005$, the time step is set to be
$\Delta t=1/128,1/256,1/1024,1/2048$. These time steps are the largest
ones that we can take to keep the stability of the explicit scheme.
We compute the solution up to $T=2$. The result is shown in Table
\ref{2nd unsteady com time}.
\begin{table}
\begin{center}
\begin{tabular}{|c||c|c|c|c|c|c|}\hline
 $N$&\multicolumn{2}{c|}{$\mu=0.05$}& \multicolumn{2}{c|}{$\mu=0.01$}
&\multicolumn{2}{c|}{$\mu=0.005$}\\ \cline{2-7} & explicit &
semi-implicit & explicit & semi-implicit & explicit & semi-implicit
\\ \hline
 64 & 7 & 0.8 & 7 & 0.8& 7 & 0.8 \\ \hline
128 & 37 & 1.6 & 38 & 1.6& 38 & 1.6  \\ \hline
256 & 249 & 4.4& 504 & 4.6& 506 & 4.5   \\
\hline
512 & 3088 & 24  & 6182 & 25& 6200 & 25  \\
\hline
\end{tabular}
\end{center}
\caption{Execution time for each computation in seconds.
Here ``explicit'' stands for the second order explicit scheme
and ``semi-implicit'' the second order semi-implicit scheme.}
\label{2nd unsteady com time}
\end{table}
Again, we observe the same qualitative behavior as the first order
schemes we reported earlier.

\section{Concluding Remarks}
In this paper, we have developed several efficient semi-implicit
immersed boundary methods for solving the immersed boundary problem
for the steady and unsteady Stokes equations. The immersed boundary
method has emerged as one of the most useful numerical methods
in computing fluid structure interaction, and has found numerous
applications. But it also suffers from the severe time step
stability limitation due to the stiffness of the elastic
force. Guided by our stability analysis, we have developed
several efficient semi-implicit schemes which remove the
stiffness of the immersed boundary method. We have demonstrated
both analytically and computationally that our semi-implicit
schemes have much better stability property than the explicit
scheme. More importantly, unlike most existing implicit or
semi-implicit schemes, our semi-implicit schemes can be
implemented very efficiently. In fact, our semi-implicit scheme
of the first kind has a computational cost that is essentially the
same as that of an explicit scheme in each time step, but
with a much better stability property. The saving in the
computational cost is quite substantial. We have demonstrated
this improved stability for a range of parameters and
numerical resolutions. Our computational results show
that the larger the spatial resolution is, the bigger the
computational saving our semi-implicit schemes can offer.
Thus the semi-implicit schemes we develop in this paper provide
an effective alternative discretization to the explicit method.

One of the essential steps in developing our semi-implicit
schemes is to obtain an unconditionally stable semi-implicit
discretization of the immersed boundary problem. This provides
us with a building block to construct our efficient
semi-implicit schemes. There are two important observations in
constructing the unconditionally stable semi-implicit discretization.
The first one is that we need to preserve certain important
solution structures at the discrete level. Specifically, we need
to ensure that the discrete velocity field is divergence free,
and the discrete spreading and interpolation operators are adjoint.
Another essential step is to decouple the stiffness of the elastic
force from the fluid flow in some appropriate way. This is difficult
to achieve if we use the Cartesian coordinate. But it becomes
easier if we use the arclength and tangent angle formulation
to describe the dynamics of the immersed interface as was done
in \cite{HLS94}. On the other hand, as we demonstrated in this
paper, the unconditionally stable semi-implicit scheme is
still very expensive to implement, and the saving over the
expicit scheme is rather limited.

Based on this unconditionally stable semi-implicit discretization,
we have developed several efficient schemes for both the steady
and the unsteady Stokes flows. By applying the Small Scale
Decomposition to the unconditionally stable semi-implicit time
discretization and further simplifying the leading order singular
kernel, we obtain our semi-implicit scheme.
The advantage of this semi-implicit scheme is that the leading
order term can be expressed as a convolution operator, which can be
evaluated explicitly using the Fourier transformation. This allows us
to solve for the implicit solution explicitly in the spectral space,
which offers substantial computational saving over the explicit scheme.

It is a natural step to extend the the semi-implicit schemes
developed for the unsteady Stokes equations to the Navier-Stokes
equations. The discretization of the Navier-Stokes equations
shares many similar properties as the unsteady Stokes equations
if we treat the convection term explicitly. We have performed
a number of numerical experiments to test the stability and the
robustness of our semi-implicit immersed boundary methods for the
Navier-Stokes equations. The results are qualitatively similar to
those for the unsteady Stokes equations which we have presented
in this paper. These results will be reported in a subsequent paper.

\vspace{0.2in} \noindent {\bf Acknowledgments.} We would like to
thank Profs. Charles Peskin and Hector Ceniceros for a number of
stimulating discussions
on the Immersed Boundary method. The research was in part supported
by DOE under the DOE grant
DE-FG02-06ER25727 and by NSF under the NSF FRG grant DMS-0353838,
ITR grant ACI-0204932, and DMS-0713670.

\appendix
\section{The semi-implicit scheme of the second kind}

In this appendix, we will derive the semi-implicit method
of the second kind in more detail. As we mentioned before,
the small scale decomposition only captures the leading order
contribution from the high frequency components, which
can not remove the stiffness induced by $S_b$ and $\mu$
completely. The coefficients $S_b$ and $\mu$ can still affect
the time stability through the low frequency components of the
solution, which comes from the lower order term of the right
hand side. In order to obtain a semi-implicit discretization
with better stability property, we can incorporate the low
frequency contribution from the second term in our implicit
discretization. For the steady Stokes flow, this gives rise to
the following decomposition:
\begin{eqnarray}
\frac{\hat{s}_{\al}^{n+1}-\hat{s}_{\al}^{n}}{\Delta t}&=&-\frac{S_b}{4\mu}
|k|\hat{s}_\al^{n+1}+\mathcal{F}\left(\theta_\al^n\int
\ln |\al-\al'| (s^{n+1}_\al-1)\theta_\al^n d\al'\right)\nonumber,\\
&&+\left[\mathcal{F}\left(V^n_\alpha-\theta^n_\alpha U^n-\theta^n_\alpha\int \ln |\al-\al'| (s^{n}_\al-1)\theta^n_\al d\al'\right)+\frac{S_b}{4\mu}|k|\hat{s}_\al^n\right],\\
\frac{\hat{\phi}^{n+1}-\hat{\phi}^{n}}{\Delta t}&=&-\frac{S_b}{4\mu}\gamma|k|\hat{\phi}^{n+1}+\mathcal{F}\left(\frac{1}{s_\al^{n+1}}V^n\theta^{n+1}_\al\right)+\left[\mathcal{F}\left(\frac{U^n_\alpha}{s_\alpha^{n+1}}\right)+\frac{S_b}{4\mu}\gamma|k|\hat{\phi}^n\right],
\end{eqnarray}
where $\D\gamma=\max_\al\left(1-\frac{1}{s_\al}\right)$.
By replacing the continuous derivative by the discrete derivative, and
discretizing the continuous integral by the trapezoidal rule, we
obtain our second semi-implicit scheme. We call this semi-implicit
scheme the semi-implicit scheme of the second kind.
Near equilibrium, we can show that the semi-implicit scheme of the second
kind is unconditionally stable. Moreover, the stability property is
independent of the meshsize, elastic coefficient $S_b$ and
viscosity coefficient $\mu$. Our numerical study also confirms this.
The trade-off is that we need to solve a linear system at each time
step to obtain the implicit solution at $t^{n+1}$.

Similarly, in the case of unsteady Stokes flow, we can also include the second
term of the right hand side in the leading order term to derive a scheme
with better stability property. In this case, the leading order term becomes:
\begin{eqnarray}
\label{new unsteady leading s}
&&\widehat{T}(s_\al^{n+1})=-\frac{S_b\Delta t}{2\left(\D\min_\al s^n_\al\right)^2}\left(\frac{\left(\lambda\D\min_\al s_\al^n\right)^2k^2+k^4}{\sqrt{\left(\lambda\D\min_\al s_\al^n\right)^2+k^2}}-|k|^3\right)\hat{s}^{n+1}_\al \nonumber\\
&&-\frac{S_b\Delta t}{2\pi}
\mathcal{F}\left(\frac{\theta^n_\alpha}{\left(s^n_\alpha\right)^2} \int \left(K_0(\lambda s^n_\alpha|\alpha-\alpha'|)-\ln(\alpha-\alpha')\right)\left((s^{n+1}_{\alpha}-1)\theta^n_{\alpha'}\right)_{\alpha'\alpha'}d\alpha'\right),\\
\label{new unsteady leading t}
&&\widehat{S}(\theta^{n+1})=-\frac{S_b\Delta t\D\max_\al\left(s^n_\al-1\right)}{2\left(\D\min_\al s^n_\al\right)^2}\left(|k|^3-\frac{k^4}{\sqrt{\left(\lambda\D\min_\al s_\al^n\right)^2+k^2}}\right)\hat{\theta}^{n+1}\nonumber \\
&& +\mathcal{F}\left(\frac{1}{s_\al^{n+1}}V^{n+1}\theta_\al^{n+1}\right) ,
\end{eqnarray}
where the derivative will be discretized by the spectral method and
the integration will be discretized by the trapezoidal rule.
We call the above scheme the semi-implicit scheme of the second kind for the
unsteady Stokes flow. Near equilibrium stability analysis suggests that
the semi-implicit scheme of the second kind is unconditionally stable.

\section{Derivation of the semi-implicit scheme (\ref{velocity ustar})-(\ref{Ustar}) }

In this appendix, we will derive the semi-implicit scheme
(\ref{velocity ustar})-(\ref{Ustar}).
We define the operator $\mathcal{G}(s_\al; \B{u}^n, \theta^n,
\B{X}^n): s_\al \rightarrow \B{u}$ by the following equations:
\begin{eqnarray}
\frac{\B{u}-\B{u}^n}{\Delta t}&=&-\nabla_h
p+\mu \nabla_h^2 \B{u}+ L_{h,n}(\B{F}(s_\alpha,\theta^{n})),\\
\nabla_h^2 p &= & \nabla_h \cdot
L_{h,n}(\B{F}(s_\alpha,\theta^{n})).\\
\end{eqnarray}
Given $s_\al$, we obtain $\B{u}$ by solving above equations. From
the definition of operator $\mathcal{G}$, we have
\begin{eqnarray}
\B{u}^{n+1}=\mathcal{G}(s_\al^{n+1};\B{u}^n, \theta^n,
\B{X}^n),\nonumber\\
\B{u}^{*,n+1}=\mathcal{G}(s_\al^{n};\B{u}^n, \theta^n,
\B{X}^n),\nonumber
\end{eqnarray}
where $\B{u}^{*,n+1}$ is calculated from equations \myref{velocity
ustar}-\myref{pstar}.

Then the equation of $s_\al$ can be rewritten as
\begin{eqnarray}
\frac{s^{n+1}_\al-s^{n}_\al}{\Delta t}&=&D_{\Delta
\al}V^{n+1}-D_{\Delta \al}\theta^n U^{n+1}\nonumber\\
&=&D_{\Delta
\al}\left(L^*_{h,n}(\B{u}^{n+1})\cdot \Bt^n\right)-D_{\Delta
\al}\theta^n \left(L^*_{h,n}(\B{u}^{n+1}\cdot \B{n}^n)\right)\nonumber\\
&=&D_{\Delta \al}\left(L^*_{h,n}\left(\mathcal{G}(s_\al^{n+1};\B{u}^n,
\theta^n, \B{X}^n)\right)\cdot \Bt^n\right)\nonumber\\
&&-D_{\Delta \al}\theta^n
\left(L^*_{h,n}\left(\mathcal{G}(s_\al^{n+1};\B{u}^n, \theta^n,
\B{X}^n)\right)\cdot \B{n}^n\right)\nonumber\\
&\equiv &\mathcal{R}(s_\al^{n+1};\B{u}^n, \theta^n, \B{X}^n).
\end{eqnarray}
Using the small scale decomposition, we get
\begin{equation}
\mathcal{R}(s_\al^{n+1};\B{u}^n, \theta^n, \B{X}^n)\sim
T(s_\al^{n+1}).
\end{equation}
By treating the leading order term implicitly, we obtain our
semi-implicit method as follows:
\begin{eqnarray}
\frac{s^{n+1}_\al-s^{n}_\al}{\Delta
t}&=&T(s_\al^{n+1})+\left(\mathcal{R}(s_\al^{n};\B{u}^n, \theta^n,
\B{X}^n)-T(s_\al^{n})\right)\nonumber\\
&=&T(s_\al^{n+1})+D_{\Delta
\al}\left(L^*_{h,n}\left(\mathcal{G}(s_\al^{n};\B{u}^n,
\theta^n, \B{X}^n)\right)\cdot \Bt^n\right)\nonumber\\
&&-D_{\Delta \al}\theta^n
\left(L^*_{h,n}\left(\mathcal{G}(s_\al^{n};\B{u}^n, \theta^n,
\B{X}^n)\right)\cdot \B{n}^n\right)-T(s_\al^{n})\nonumber\\
&=&T(s_\al^{n+1})+D_{\Delta \al}\left(L^*_{h,n}(\B{u}^{*,n+1})\cdot
\Bt^n\right)-D_{\Delta \al}\theta^n \left(L^*_{h,n}(\B{u}^{*,n+1})\cdot
\B{n}^n\right)\nonumber\\
&&-T(s_\al^{n})\nonumber\\
&=&T(s_\al^{n+1})+\left(D_{\Delta \al}V^{*,n+1}-D_{\Delta
\al}\theta^n U^{*,n+1}-T(s_\al^{n})\right).
\end{eqnarray}
This is exactly our semi-implicit scheme \myref{semi implicit
sa}.

In the steady Stokes case, we can define the operator
$\mathcal{G}(s_\al; \B{u}^n, \theta^n,
\B{X}^n): s_\al \rightarrow \B{u}$ similarly:
\begin{eqnarray}
0&=&-\nabla_h p
+\mu \nabla_h^2 \B{u}+L_{h,n}(\B{F}(s_\alpha,\theta^{n})),\\
\nabla_h^2 p&=&\nabla_h\cdot L_{h,n}(\B{F}(s_\alpha,\theta^{n})).
\end{eqnarray}
In this case, we have
\begin{eqnarray}
\mathcal{G}(s_\al^{n};\B{u}^n, \theta^n,
\B{X}^n)=\B{u}^{n} .\nonumber
\end{eqnarray}
Thus, the semi-implicit scheme becomes
\begin{eqnarray}
\frac{s^{n+1}_\al-s^{n}_\al}{\Delta
t}=T(s_\al^{n+1})+\left(D_{\Delta \al}V^{n}-D_{\Delta
\al}\theta^n U^{n}-T(s_\al^{n})\right).
\end{eqnarray}

\section{The derivation of the Fourier transform of $K_0$}
In this appendix, we derive the the Fourier transform of $K_0$,
which is given in (\ref{fft of K0}).
By the definition of the Fourier transform, we have
\begin{eqnarray}
&&\mathcal{F}\left(\frac{1}{\pi}\int_{-\infty}^{+\infty}K_0(\beta |\al-\al'|)f(\al')d\al'\right)\nonumber\\
&=&\frac{1}{\pi}\int_{-\infty}^{+\infty}\left(\int_{-\infty}^{+\infty}K_0(\beta|\al-\al'|)f(\al')d\al'\right)e^{-ik\al}d\al\nonumber\\
&=&\frac{1}{\pi}\int_{-\infty}^{+\infty}\int_{-\infty}^{+\infty}K_0(\beta|\al-\al'|)f(\al')e^{-ik\al}d\al'd\al\nonumber\\
&=&\frac{1}{\pi}\int_{-\infty}^{+\infty}\int_{-\infty}^{+\infty}K_0(\beta|\al-\al'|)f(\al')e^{-ik(\al-\al')}e^{-ik\al'}d\al'd(\al-\al')\nonumber\\
&=&\frac{1}{\pi}\int_{-\infty}^{+\infty}\left(\int_{-\infty}^{+\infty}K_0(\beta|\al-\al'|)e^{-ik(\al-\al')}d(\al-\al')\right)f(\al')e^{-ik\al'}d\al'\nonumber\\
&=&\int_{-\infty}^{+\infty}\frac{f(\al')}{\sqrt{\beta^2+k^2}}e^{-ik\al'}d\al'
=\frac{\widehat{f}(k)}{\sqrt{\beta^2+k^2}} ,
\end{eqnarray}
where
$\D\mathcal{F}(f(\al))(k)=\D\int_{-\infty}^{+\infty}f(\al)e^{ik\al}d\al$
is the Fourier transform. In the calculation above, we have used the
expression of the Bessel function (p 376, \cite{AS72})
\begin{eqnarray}
K_0(x)=\int_{0}^{+\infty}\frac{\cos(tx)}{\sqrt{1+t^2}}dt,
\end{eqnarray}
and the identity below:
\begin{eqnarray}
\int_{-\infty}^{+\infty}K_0(\beta |x|)e^{-ikx}dx
&=&\int_{-\infty}^{+\infty}\int_{0}^{+\infty}\frac{\cos(\beta tx)}{\sqrt{1+t^2}}e^{-ikx}dtdx\nonumber\\
&=&\frac{1}{2}\int_{-\infty}^{+\infty}\int_{-\infty}^{+\infty}\frac{\cos(\beta tx)}{\sqrt{1+t^2}}e^{-ikx}dtdx\nonumber\\
&=&\frac{1}{2}\int_{-\infty}^{+\infty}\int_{-\infty}^{+\infty}\frac{e^{i\beta tx}}{\sqrt{1+t^2}}e^{-ikx}dtdx\nonumber\\
&=&\frac{1}{2}\int_{-\infty}^{+\infty}\int_{-\infty}^{+\infty}\frac{e^{i(\beta t-k)x}}{\sqrt{1+t^2}}dtdx\nonumber\\
&=&\pi\int_{-\infty}^{+\infty}\frac{\delta(\beta
t-k)}{\sqrt{1+t^2}}dt =\frac{\pi}{\sqrt{\beta^2+k^2}}.
\end{eqnarray}
This proves (\ref{fft of K0}).

\end{document}